\documentclass{IEEEtran}

\usepackage{xcolor}
\usepackage{amsmath,amssymb,amsfonts}
\usepackage{amsthm}
\usepackage{graphicx}
\usepackage{booktabs}

\newcounter{defCounter}
\newcounter{LemCounter}
\newcounter{PropCounter}

\newcommand{\frm}[1]{\langle #1\rangle}

\newtheorem{theorem}{Theorem}

\newtheorem{case}{Case}

\newtheorem{definition}[defCounter]{Definition}

\newtheorem{lemma}[LemCounter]{Lemma}

\newtheorem{remark}{Remark}
\newtheorem{fact}{Fact}

\newtheorem{assumption}{Assumption}
\newtheorem{property}[PropCounter]{Property}

\newtheorem*{notation}{Notation}

\theoremstyle{definition}
\newtheorem{exx}{Example}
\newenvironment{example}{\begin{exx}}{\hfill \hspace*{1pt} \hfill $\star$ \end{exx}}

\DeclareMathOperator*{\diag}{diag}
\DeclareMathOperator{\atantwo}{arctan2}
\newcommand{\ind}[1]{\operatorname{Ind}({#1})}
\newcommand{\C}[1]{$\mathbf{C{#1}}$}

\title{Analysis of Indistinguishable Trajectories of a Nonholonomic
Vehicle Subject to Range Measurements}

\author{Francesco Riz, Luigi Palopoli, Daniele Fontanelli
\thanks{
F. Riz and L. Palopoli are with the Department of Information Engineering
and Computer Science, University of Trento, Italy, e-mail:
\texttt{\{francesco.riz, luigi.palopoli\}@unitn.it}.
}
\thanks{
D. Fontanelli is with the Department of Industrial Engineering,
University of Trento, Italy, e-mail: \texttt{daniele.fontanelli@unitn.it}
}
}

\begin{document}
    \maketitle
    \begin{abstract}
      We propose a \emph{global} constructibility analysis for a
      vehicle moving on a planar surface. Assuming that the vehicle
      follows a trajectory that can be uniquely identified by the
      sequence of control inputs and by some intermittent ranging
      measurements from known points in the environment, we can model
      the trajectory as a rigid body subject to rotation and
      translation in the plane. This way, the localisation problem can
      be reduced to finding the conditions for the existence of a
      unique roto-translation of the trajectory from a known reference
      frame to the world reference frame, given the collected
      measurements. As discussed in this paper, such conditions can be
      expressed in terms of the shape of the trajectory, of the layout
      of the ranging sensors, and of the numbers of measurements
      collected from each of them. The approach applies to a large
      class of kinematic models. Focusing on the special case of
      unicycle kinematics, we provide additional \emph{local}
      constructibility results.
    \end{abstract}

        

    \section{Introduction}
\label{sec:introduction}

Mobile robots are increasingly popular in many real--life
applications, where they are required to plan and execute complex
trajectories (e.g., to avoid humans and complex dynamic obstacles).
Inevitably, these operations require an accurate \textit{localisation}
of the robot in the environment. The most common strategies to solve
this problem are based on a combination of odometry, which provides
information on the manoeuvres executed by the robot, and absolute
measurements collected through exteroceptive sensors.  In the class of
exteroceptive sensors used in modern robotics fall those based on
radio frequency signals~\cite{MagnagoCPPF19iros}, which are attracting
interest and consideration due to their robustness, flexibility and
the relatively low cost. Such sensing systems  rely on some features
of the sensed signal such as the sensed power or the time--of--flight,
are based on technologies such as Wi--Fi~\cite{Chen2013survey}, Radio
Frequency Identification (RFID)~\cite{NazemzadehMFMP15} or Ultra Wide
Band (UWB)~\cite{Cheok10} signals, and generate measurements that are
functionally related to the distance between a fixed-frame point and
the vehicle, exactly as a LiDAR~\cite{Gallant16} or a
radar~\cite{PosnerB20} would do. Other types of sensors, which can be
used in an indoor environment, are based mainly on visual information
collected by cameras~\cite{belo2013complete}, and in most of the cases
produce \textit{bearing} measurements.

In this paper, we consider a vehicle moving in an environment,
instrumented with ranging sensors. The vehicle is initially unaware of
its position and orientation in the space. In this setting, we analyse
the properties of the environment, i.e. number and layout of the
sensors, and the ``shape'' of the trajectory, i.e. the sequence of
manoeuvres, that allow the vehicle to localise itself.

\textbf{Related work:} The problem considered in this paper is often
addressed in the literature from the viewpoint of the
\textit{observability analysis}. Very often the most important results
rely on the \textit{Observability Rank Condition} (ORC), i.e. a system
is observable only if the \textit{Observability Matrix} has full rank.
By using ORC, Belo et al.~\cite{belo2013complete} carry out a complete
observability analysis of a system composed of moving vehicles
(targets) and moving cameras (sensors), collecting planar bearing
measurements. Other researchers extend this type of analysis to
multiple fixed-frame landmarks in a 3D
environment~\cite{sert2012localizability}, use the same tools to
dynamically find the optimal control
strategy~\cite{mariottini2005vision}, or implement estimation filters
based on a pipeline that starts with visual information collected from
cameras, extracts features, and localises the vehicle in the
environment~\cite{wu2020robust}.  Delaune et
al.~\cite{delaune2021range} show that, in some particular cases,
bearing measurements are not sufficient to reconstruct the trajectory
followed by the vehicle, and they propose to integrate also the
information coming from range sensors. Based on the Observability
Matrix, Martinelli et al.~\cite{martinelli2017unicycle} analyse the
observability of a vehicle subject to a single measurement, be it
bearing or range. Single landmarks measuring their distance from the
target vehicle have been considered in~\cite{de2017underwater}
and~\cite{quenzer2014observability}, where the authors build the
Observability Matrix and the Observability Gramian, respectively, to
quantify the observability of the system. Fernando et
al.~\cite{fernando2021toward} analyse how the number of ranging
sensors affects the observability of a Micro Aerial Vehicle and show
that the observability properties of the system heavily depend on the
manoeuvres executed by the vehicle. Magnago et
al.~\cite{MagnagoPBTMNMF20tim} use RFID tags and show that a suitably
designed Unscented Kalman Filter converges only in presence of at
least $3$ tags. Other researchers use only ranging information to
estimate the state of the
system~\cite{fontanelli2021uncertainty,araki2019range}. A noteworthy
area of research is \emph{active sensing}, i.e., designing control
strategies that maximise some observability metrics. This technique
may be applied to a moving sensor, finding the trajectory that
optimises the observability of a moving
target~\cite{hung2020range,coleman2021observability,
mandic2016mobile}, or to the moving vehicle itself which senses some
fixed-frame
sensors~\cite{cedervall2007nonlinear,salaris2019online,napolitano2021gramian}.

Restricting to authors who directly addressed global observability
properties for vehicles, Bayat et al.~\cite{bayat2015range} analyse
the observability properties of an underwater vehicle, modelled as a
2D single integrator with known heading, subject to range
measurements. Similar results have been obtained
in~\cite{de2017underwater}, where the analysed vehicle is represented
by a double integrator. The authors define the displacement as the
double integration of the control inputs of the vehicle, and analyse
the equivalent augmented linear time varying system by means of the
observability Gramian. With a similar procedure, Hung et
al.~\cite{hung2020range} analyse an underwater robot modelled as a 3D
single integrator vehicle moving through unknown ocean currents. They
exploit the displacement of the vehicle, which is known via control,
to define an augmented state vector and on output equation with linear
form $\bar y(t) = C(t) z(t_0)$, where $\bar y(t)$ is the augmented
output, which is known and based on the collected measurements, while
$z(t_0)$ is the unknown initial augmented state.  Observability is
tested by checking the singularity of matrix $C(t)$ over the time
interval when measurements are collected.  A common requirement of the
research papers mentioned above is that the heading of the vehicle is
assumed known, and so is the information of the vehicle displacement
in the world reference frame. In this paper, we consider a scenario
where: 1. measurements are intermittent, 2. vehicles belong to a
general class, including the unicycle--like vehicles, for which the
displacement in the world reference frame depends on the sequence of
control inputs and on the initial (unknown) orientation, 3. the
heading is not measurable.  As a result, the displacement in the world
frame is not known, which makes the results mentioned above
inapplicable.

\textbf{Paper contributions:}
The main part of the technical literature described so far uses the
Observability Matrix or the Observability Gramian as tools to quantify
the observability of a system. However, since these tools are based on
the linearisation of the dynamics of the system or of the output
function associated with the measurements collected by the sensors,
they produce local results, which are associated with the notion of
\textit{weak observability}. By definition, a weakly constructible (or
a weakly observable) system can reconstruct its state along its
trajectory as long as it is provided with some \textit{a priori}
information on its state at a certain time instant, such as a
sufficiently narrow set that includes the state itself. In light of
this consideration, we analyse the setting where no \textit{a priori}
information is given, thus referring to ``global
observability/constructibility properties''. In our past
work~\cite{palopoli2020global}, we have analysed global observability
properties, based on the concept of \textit{indistinguishable} states,
in presence of ranging sensors with unbounded sensing range. As a
follow--up~\cite{riz2022local}, we have proposed a sufficient
condition for attaining global observability in the case of bounded
sensing range for a unicycle kinematic model. In this paper we
abstract the dynamics of the system by considering a finite number of
points that can be regarded as roto-translations of a given sequence
of points in a known reference frame. With this consideration, our aim
is twofold: first we extend the global constructibility analysis (more
formally the \textit{$u$-constructibility analysis}) with intermittent
measurements, and provide both sufficient and necessary conditions to
achieve global constructibility. Considering intermittent measurement
is key in defining the scope and the contribution of this paper.  From
the modelling point of view, it allows us to consider real--life
sensors with bounded sensing range, From the analysis point of view,
it enables us to carry out the global analysis by simple geometric
intuitions. Secondly, we consider a particular nonholonomic vehicle
(unicycle kinematic model) and analyse its \emph{local}
constructibility properties in the same scenario. We show that,
despite the logical intuition, the global property does not always
imply local constructibility, and discuss some degenerated cases where
this implication does not hold true.

The paper is organised as follows: in
Section~\ref{sec:problem_statement} we introduce the dynamical and
measurement model, and an abstraction of the trajectory allowing us to
interpret constructibility properties from a geometrical perspective.
With this assumption, Section~\ref{sec:single_anchor}
and~\ref{sec:more_anchors} analyse the conditions on the shape of the
trajectory, on the layout of the sensors, and on the number of
measurements and their distribution ensuring constructibility. In
Section~\ref{sec:local_constructibility}, we present a local
constructibility analysis based on the Constructibility Gramian and,
in Section~\ref{sec:conclusions}, we draw the conclusions and claim
some further research directions.

    \section{Background and Problem Formulation}
\label{sec:problem_statement}

Let us consider a generic continuous-time nonlinear system in its
state space representation
\begin{equation}
    \label{eq:CT_system_dynamics}
    \dot q(t) = f(q(t),u(t)),
\end{equation}
where $q\in \mathbb{R}^n$ is the state of the system, while $u\in
\mathbb{R}^m$ denotes its control inputs. We assume that the nonlinear
system represents the dynamics of a vehicle, moving on a planar
surface, and thus a portion of the state vector $q(t)$ denotes the
Cartesian position $P(t) = [x(t),y(t)]^\top$ of the vehicle in a
reference frame $\frm W$ on the plane $X_w\times Y_w$. We denote by
$\frm V$ a reference frame where the initial conditions $q_V(0)$ of
the vehicle are arbitrarily set to $0$, i.e. the reference frame is
centred on the initial position of the vehicle. In $\frm V$, the
position of the vehicle at time $t$ is represented by $P_V(t) =
[x_V(t), y_V(t)]^\top$ and can be reconstructed by using the control
input history $u(s),\,s\in[0, t]$. We will use the following
Property~\ref{property:dynamical_systems}, which is directly derived
from the definition of the rotation matrix $R_\phi =
\left[\begin{smallmatrix} \cos\phi & -\sin\phi \\ \sin\phi & \cos\phi
\end{smallmatrix}\right]$.

\begin{property}
\label{property:dynamical_systems}
Given the position of the vehicle $P_V(t),\,\forall t \in [t_0,t_f]$,
there exists a unique triplet $(\Delta x, \Delta y,\phi)$ such that
\begin{equation*}
  P(t) = R_\phi P_V(t) + \begin{bmatrix}
    \Delta x \\ \Delta y
    \end{bmatrix},\quad\forall t \in [t_0,t_f] .
  \end{equation*}
\end{property}
From a geometrical point of view, Property 1 states that the path
followed by the vehicle in any reference frame is a roto-translation
of the path travelled in its local reference frame, i.e. the vehicle
can reconstruct the ``shape'' of its own trajectory by dead reckoning.
By Property~\ref{property:dynamical_systems}, we may simplify the
dynamics~\eqref{eq:CT_system_dynamics} of the vehicle and consider the
path followed by the vehicle as a rigid body on the $X_w\times Y_w$
plane.

The environment is instrumented with a set of sparsely deployed
ranging sensors, referred to as {\em anchors}. The $i$-th anchor is
located at coordinates $B_i = [X_i,Y_i]^\top,\,i=1,\dots ,p$, and the
vehicle collects the ranging measurement $\|B_i - P(t)\|$.
\begin{assumption}[Intermittent measurements]
  Measurements are collected at known sampling instants $t_k$, with
  $t_{k+1} > t_k$.
\end{assumption}
In ranging scenarios, the vehicle is usually equipped with
\emph{active} devices that sense other devices (active or passive)
deployed in the environment, and thus assuming \emph{known} sampling
instants is not a demanding condition. For example, this condition
holds for radio-frequency ranging measurements (such as Ultra-Wide
Band anchors~\cite{MagnagoCPPF19iros} or Radio Frequency
IDentification tags~\cite{MagnagoPBTMNMF20tim}) and for camera based
solutions~\cite{NazemzadehMFMP15}.
The output $z_k$ is given by the measurements collected by the
anchors, i.e., the output equation is the following:
\begin{equation}
  \label{eq:DT_output_equation}
  z_k = \|P_k - B_{i_k}\|,
\end{equation}
where $P_k = P(t_k)$ is the position of the vehicle at time $t = t_k$,
while the index $i_k \in \{1,\dots,p\}$ defines the anchor that the
vehicle measures its distance from. For the sake of clarity, the
collected distance will be denoted by $\rho_{k,i}$ when the second
index $i$ is not clear from the context. Measurements are
intermittent; therefore at time $t_k$, only one ranging measurement
$\rho_{k,i}$ is available. The case when multiple measurements can be
collected at once has been already solved
in~\cite{palopoli2020global}.

We assume full knowledge of the time instants $t_k$ when the
measurements are taken and of the input sequence $u(s),\,s\in[0,
t_k]$, which allows us to reconstruct the sequence  of positions
$P_V(t_k)$ of the vehicle in $\frm V$. Therefore, instead of
considering the entire paths $P(t)$ and $P_V(t)$, we focus only on the
locations where the ranging measurements are collected:
\begin{equation*}
    \mathcal{P}_k (\Delta x, \Delta y, \phi) = R_\phi P_V(t_k) + \begin{bmatrix}
      \Delta x \\ \Delta y
    \end{bmatrix},
\end{equation*}
for $k=0,\dots,N_m-1$, with $N_m$ being the total number of
measurements. Given two points $\mathcal{P}_l$ and $\mathcal{P}_m$, we
define by $\mathcal{S}_{l,m}$ the segment given by their convex
combination, with length $\|\mathcal{S}_{l,m}\|$. We can restrict our
study to an abstract trajectory $\mathcal{T}$, defined as the union of
all the segments connecting two consecutive positions $\mathcal{P}_k$
of the vehicle, thus
\begin{equation}
    \label{eq:trajectory_T}
    \mathcal{T} = \bigcup_{k=0}^{N_m-1} \mathcal{S}_{k,k+1},
\end{equation}
which can be regarded as a rigid body.
\begin{remark}
  The abstract trajectory $\mathcal{T}$ does not coincide with the
  actual trajectory $P(t)$, but contains all the features that are
  needed in the following discussion: the sequence of measurements,
  the distance and the total change in the orientation between any two
  measurements.
\end{remark}

In light of the definition of $\mathcal{T}$, we want to find the
conditions on the position of the anchors in $\frm W$ and on the
trajectory $\mathcal{T}$ such that it is possible to find a
roto-translation such that the points $\mathcal{P}_k$ are compliant
with the measurements collected from the anchors. To this aim, we need
to introduce the concepts of \textit{constructibility} and
\textit{backward indistinguishability} of the states of a nonlinear
system. For the sake of generality, in the following definitions,
adapted from~\cite{bayat2015range}, we consider a plant with a
continuous-time dynamics~\eqref{eq:CT_system_dynamics} and the general
version of the discrete-time output
equation~\eqref{eq:DT_output_equation}, i.e.
\begin{equation}
    \label{eq:CT_DT}
    \dot q(t) = f(q(t),u(t)), \qquad
    z_{k}     = h(q(t_k)).
\end{equation}
We consider the dynamical system to evolve between an initial time
instant $t_0$ (i.e. $k=0$) and a final time $t_f$, with $k=k_f$. Given
the hybrid nature of~\eqref{eq:CT_DT}, we will use both $k$ and $t$ to
denote the time, with the implicit assumption that by the time instant
$k$ we refer to time $t_k$.

We deal with the notion of \emph{constructibility}, defined as the
ability to reconstruct the state $q_f$ of the system at the final time
instant $t_f$. Intuitively, constructibility amounts to reconstructing
the current state $q_k$ given the past history of inputs and outputs.
In view of Property~\ref{property:dynamical_systems}, the problem of
estimating the final state is equivalent to estimating the initial
state $q_0$, which is the well-known concept of
\textit{observability}. However, the performance of an estimation
filter, i.e. the uncertainty related to the state estimate based on
the previous history of motions and measurements, is not directly
associated with the concept of observability, but rather to the notion
of \emph{constructibility} (for further details, the reader is
referred to~\cite{salaris2019online}). To analyse formally the concept
of constructibility, both from a local and from a global perspective,
we introduce the definition of \textit{backward indistinguishability},
instead of the notion of \emph{indistinguishability} considered for
more common observability analyses~\cite{bayat2015range}.

\begin{definition}
  \label{def:u_constr}
  Given the dynamical system~\eqref{eq:CT_DT}, a time interval
  $T=[t_0,t_f]$, and an admissible control input function
  $u^\star(t),\,t\in T$, two final states $q_f$ and $\bar q_f$ are
  said $\mathbf{u^\star}$--\textbf{backward indistinguishable} on $T$,
  if for the input $u^\star(t),\,t\in T$, the output sequences $z_k$
  and $\bar z_k,\,k=0,\dots,k_f$ of the trajectories satisfying the
  final conditions $q_f$, $\bar q_f$, are identical. Moreover, we
  define $\mathcal{I}_{(b)}^{u^\star}(q_f)$ as the set of all the
  final conditions that are $u^\star$--backward indistinguishable from
  $q_f$ on $T$.
\end{definition}
Since we assume full knowledge of the control input, the shape of the
trajectory $\mathcal{T}$  is known in its turn. Hence, we will focus
on the concept of $u$-backward indistinguishability.  With a slight
abuse of definition, we will refer to \textit{indistinguishable}
trajectories as trajectories generated by a known control input  and
by two $u$--\textit{backward indistinguishable} final conditions.

We now introduce further definitions on \textit{constructibility} that
will be useful for the local analysis carried out in
Section~\ref{sec:local_constructibility}.

\begin{definition}
\label{def:weakly_constructible}
Given the interval $T=[t_0,t_f]$, and the control input
$u^\star(t),\,t\in T$, the system~\eqref{eq:CT_DT} is said
$\mathbf{u^\star}$--\textbf{constructible at} $\mathbf{q_f}$ on $T$,
if $\mathcal{I}_{(b^)}^{u^\star}(q_f) = \{q_f\}$, and  is said
$\mathbf{u^\star}$--\textbf{weakly constructible at} $\mathbf{q_f}$ if
$q_f$ is an isolated point of $\mathcal{I}_{(b)}^{u^\star}(q_f)$.
Moreover, a system is said to be $\mathbf{u^\star}$\textbf{--(weakly)
constructible} if it is $u^\star$--(weakly) constructible at
\emph{any} $q_f$.
\end{definition}

In the local analysis, associated with the concept of \textit{weak
constructibility}, we will refer to a \textit{weakly constructible
trajectory} as a trajectory, defined by a control sequence $u^\star$,
such that the system is $u^\star$-\textit{weakly constructible}. By
Definition~\ref{def:weakly_constructible}, the set of \emph{globally
constructible} trajectories is a subset of the \emph{weakly
constructible} trajectories, since a unique point in a set is always
isolated. Therefore, if a system is \emph{weakly unconstructible},
i.e. the Observability Rank Condition is \emph{not} satisfied, the
system is never \emph{constructible}. 

\begin{assumption}
\label{assum:initial}
We assume that system~\eqref{eq:CT_DT} with the continuous--time
position $P(t)$ as output is \emph{constructible}, i.e. we can
reconstruct its final state $q_f$ given the history of inputs and
outputs over a given time interval.
\end{assumption}
\begin{remark}
Assumption~\ref{assum:initial} holds true for nonholonomic systems
modelled through kinematic models (e.g. unicycle--like vehicles and
car--trailer vehicles). Moreover, vehicles represented by dynamical
models meet this assumption as soon as they can rely on additional
sensors measuring \emph{at least} their velocities.
\end{remark}

We can now link the notion of constructibility to the existence of a
roto-translation that, when applied to $\mathcal{T}$, produces a set
of points compliant with the measurements.

\begin{lemma}
  \label{lem:equivalent}
 The system is $u^\star$-constructible if there exists a unique
  roto-translation $(\Delta x, \Delta y, \phi)$ of $\mathcal{T}$
  generated by $u^\star$ such that
  \begin{equation*}
    \|\mathcal{P}_k(\Delta x, \Delta y, \phi) - B_{i}\| = \rho_{k,i},
  \end{equation*}
  for each $i$ such that the measurement is available at time $k$, and
  for $k=0,\dots,k_f$.
\end{lemma}

\begin{proof}
The proof directly follows from
Property~\ref{property:dynamical_systems} and from
Assumption~\ref{assum:initial}.
\end{proof}

\begin{remark}
  By Assumption~\ref{assum:initial}, as soon as a roto-translation is
  found, we can reconstruct the entire trajectory followed by the
  vehicle and thus retrieve both the initial condition $q(t_0)$ and
  the final condition $q(t_f)$. Therefore, the system is observable if
  and only if it is constructible. By following the same arguments, a
  system is weakly observable if and only if it is weakly
  constructible.
\end{remark}

For the sake of simplicity, we introduce here a new definition to link
the number of indistinguishable trajectories to the constructibility
properties of the system.
\begin{definition}
  \label{def:ind_n}
  Given a trajectory $\mathcal{T}$, if there exist $n$
  roto-translations of $\mathcal{T}$ that are indistinguishable from
  $\mathcal{T}$ itself, we say that $\mathcal{T}$ is $\ind{n}$.
\end{definition}

Definition~\ref{def:ind_n} is associated only with the global
constructibility properties, and, since a trajectory $\mathcal{T}$ is
always indistinguishable from itself, it is impossible to have
$\ind{0}$. On the other hand, a system is constructible if and only if
it is $\ind{1}$. Moreover, in light of Definition~\ref{def:ind_n}, a
system is unconstructible when $\mathcal{T}$ is
$\operatorname{Ind}(\infty)$, i.e. when there is an infinite set of
roto-translations from $\frm V$ to $\frm W$
satisfying~\eqref{eq:DT_output_equation}.

\subsection{Problem Statement}
Given a dynamical system such that
Property~\ref{property:dynamical_systems} holds, and the sequence of
positions $P_V(t_k),\,k=0,\dots,k_f$, in the vehicle reference frame,
we want to find the conditions on $\mathcal{T}$, on the layout of the
sensors $B_i = [X_i, Y_i]^\top,\,i=1,\dots,p$, in $\frm W$, and on the
distribution of the measurements among the sensors, such that the
system is $u$-constructible (Sections~\ref{sec:single_anchor}
and~\ref{sec:more_anchors}) and $u$-weakly constructible
(Section~\ref{sec:local_constructibility}) at the final condition
$q_f$. In light of the discussion above, these problems boil down to
find whether the equations
\begin{equation*}
    \left\|\mathcal{P}_k(\Delta x, \Delta y, \phi) - B_i\right\| =
    \rho_{k,i},\,\,\, \forall k
\end{equation*}
have a unique solution, a finite number of solutions or infinitely
many solutions in the unknowns $\Delta x, \Delta y, \phi$, where the
roto-translation of $\mathcal{T}$ from $\frm V$ to $\frm W$ is
modelled by the translation vector $[\Delta x; \Delta y]^\top$ and by
the rotation angle $\phi$.
    \section{Global constructibility with a single anchor}
\label{sec:single_anchor}

We consider the trajectory $\mathcal{T}$ defined
in~\eqref{eq:trajectory_T}, and discuss how the readings of a single
anchor change depending on roto-translations of $\mathcal{T}$. In our
past work~\cite[Thm. 1]{palopoli2020global}, we have restricted our
analysis to a unicycle-like vehicle, and we have proved that a single
anchor collecting range measurements can never ensure observability of
the system state. In simple terms, we have proved that
$u$-constructibility as in Definition~\ref{def:u_constr} can never be
achieved with a single anchor. We now want to reformulate this result
in terms of the trajectory $\mathcal{T}$ and generalise the analysis
of the non-constructible subspaces of the system, depending on the
number and on the layout of the measurement points sensed by the
anchor.

Without loss of generality, we will consider one anchor located at the
origin of the reference frame $\frm W$, i.e. $B = [0,0]^\top$, and we
will focus on the first three points $\mathcal{P}_0$, $\mathcal{P}_1$
and $\mathcal{P}_2$ of $\mathcal{T}$, where the measurements occur.
\begin{theorem}
\label{thm:single_indistinguishable}
Given a vehicle for which Property~\ref{property:dynamical_systems}
holds, its trajectory $\mathcal{T}$ and the set of measurements
$\rho_k$, $k = 0,\dots, N_m-1$, collected from an anchor $B$, a
trajectory $\bar{\mathcal{T}}$ is $u$-indistinguishable from
$\mathcal{T}$ if: (1) For any $N_m$, $\bar{\mathcal{T}}$ is a rotation
of $\mathcal{T}$ about the anchor; (2) For $N_m=1$ (or $N_m>1$
coincident points $\mathcal{P}_k$), $\bar{\mathcal{T}}$ is a rotation
of $\mathcal{T}$ about the unique measurement point $\mathcal{P}_0$;
(3) For $N_m=2$ (or $N_m>2$ with collinear points $\mathcal{P}_k$),
$\bar{\mathcal{T}}$ is symmetric to $\mathcal{T}$ with respect to an
axis passing through the anchor.


  

\end{theorem}

\begin{proof}
By geometric arguments, any rotation of the trajectory about the
anchor does not change the sensor readings, and thus the system sensed
with a single anchor is always (at least) $\ind{\infty}$.  We now
analyse scenarios with increasing number of measurements collected by
the anchor.

\paragraph*{One measurement}
With one measurement, we identify a point $\mathcal{P}_0$ that is
sensed by the anchor, thus constraining the possible roto-translations
of $\mathcal{T}$ that satisfy the sensor readings. The measurement
point is compliant with the sensor reading only for a position
$\mathcal{P}_0 = [\rho_0 \cos \phi, \rho_0 \sin \phi]^\top$, for any
$\phi \in [0,2\pi)$. Therefore, the trajectory is compliant with the
measurement for any rotation of the trajectory about the anchor plus
any rotation about $\mathcal{P}_0$, i.e. the system is
$\ind{\infty\times\infty}$, unless $\rho_0 = 0$, i.e. $\mathcal P_0$
coincides with the anchor.

\paragraph*{Two measurements}
By taking the second measurement in position $\mathcal{P}_1$, provided
that the two measurements are not taken in the same point (otherwise
the previous case straightforwardly applies), we are adding a further
constraint on the position and orientation of the trajectory. Indeed,
with $\mathcal P_0 = R_\phi [\rho_0;0]^\top$ and $\mathcal P_1 =
R_\beta [\rho_1;0]^\top$, and $\|\mathcal P_1 - \mathcal P_0\| =
\|\mathcal S_{0,1}\|$, we get an explicit expression of $\beta$, which
reads as
\begin{equation}
    \label{eq:two_measurements_beta}
    \beta = \phi \pm \arccos\left(%
        \frac{\rho_1^2-\|\mathcal{S}_{0,1}\|^2-\rho_0^2}%
        {2\rho_0\|\mathcal{S}_{0,1}\|}\right) \triangleq \phi \pm \delta.
\end{equation}
This result shows that, for any rotation $\phi$ about the anchor,
there are two different points $\mathcal{P}_1^{(a)}$ and
$\mathcal{P}_1^{(b)}$, that are compliant with the manoeuvres executed
by the vehicle (i.e. $\|\mathcal{S}_{0,1}\|$) and the measurements
collected by the anchor (i.e. $\rho_0$ and $\rho_1$), hence this
setting leads to a $\ind{2\times\infty}$ system.

The geometric interpretation of~\eqref{eq:two_measurements_beta} is a
reflection about an axis passing through the anchor $B$. Indeed, any
point of a circle reflected about an axis passing through its centre
lies on the circle itself. Moreover, the geometry of the trajectory,
which is uniquely identified by the distance $\|\mathcal S_{0,1}\|$,
is preserved.

\paragraph*{Three measurements}
Let us consider the setting with two measurements presented
previously. For each of the two values of $\beta$, we can compute
explicitly the position of the third measurement point $\mathcal
P_2^{(a)}$ and $\mathcal P_2^{(b)}$, represented in
Figure~\ref{fig:two_plus_one}.
\begin{equation}
    \label{eq:two_points_distance}
    \small
    \arraycolsep=1pt
\begin{array}{ccl}
    \mathcal{P}_2^{(a)} &=& \left[\begin{array}{ccccc}
        \rho_0 &+& \|\mathcal{S}_{0,1}\| \cos( \delta) &+& \|\mathcal{S}_{1,2}\| \cos( \delta + \mu_{0,1}) \\
        && \|\mathcal{S}_{0,1}\| \sin( \delta) &+& \|\mathcal{S}_{1,2}\| \sin( \delta + \mu_{0,1})
    \end{array}\right],\\[10pt]
    \mathcal{P}_2^{(b)} &=& \left[\begin{array}{ccccc}
        \rho_0 &+& \|\mathcal{S}_{0,1}\| \cos(-\delta) &+& \|\mathcal{S}_{1,2}\| \cos(-\delta + \mu_{0,1}) \\
        && \|\mathcal{S}_{0,1}\| \sin(-\delta) &+& \|\mathcal{S}_{1,2}\| \sin(-\delta + \mu_{0,1})
    \end{array}\right],
\end{array}
\end{equation}
where $\mu_{0,1}$ is the angle between the segments
$\mathcal{S}_{0,1}$ and $\mathcal{S}_{1,2}$. For simplicity, but
without loss of generality, we have assumed $\phi = 0$
in~\eqref{eq:two_measurements_beta}. Computing the differences of the
distances of $\mathcal{P}_2^{(a)}$ and $\mathcal{P}_2^{(b)}$ from the
origin, we have
\begin{equation*}
    \|\mathcal{P}_2^{(b)}\|^2 - \|\mathcal{P}_2^{(a)} \|^2=
    4 \rho_0 \|\mathcal{S}_{1,2}\| \sin \mu_{0,1} \sin \delta.
\end{equation*}
The two distances are equal when $\mu_{0,1} = h\pi,\,h\in\mathbb Z$,
i.e. when $\mathcal{P}_0$, $\mathcal{P}_1$ and $\mathcal{P}_2$ are
collinear, or when $\delta = h\pi,\,h\in\mathbb Z$, i.e. the situation
described in Remark~\ref{rem:two_measurements_diameter_global} occurs,
hence $\mathcal{P}_2^{(a)}$ and $\mathcal{P}_2^{(b)}$ coincide.
Therefore, only trajectories rotated around the anchor are
indistinguishable, hence the problem is $\ind{\infty}$.
\end{proof}

\begin{remark}
  \label{rem:two_measurements_diameter_global}
  In the particular case when $\rho_1 = \rho_0 \pm
  \|\mathcal{S}_{0,1}\|$, i.e. the vehicle moves on the diameter of
  the circle centred on the anchor, we get $\cos\delta = \pm 1$, i.e.
  a unique feasible value for $\beta$
  in~\eqref{eq:two_measurements_beta}, hence avoiding the ambiguity
  associated with the rotation about $\mathcal{P}_0$, i.e.
  $\ind{1\times\infty} = \ind{\infty}$.
\end{remark}

With three non-collinear measurement points, we reach the maximum
amount of information that can be collected by a single anchor, and
thus we conclude that any further measurement beyond the third is no
more informative (unless all the preceding measurement points are
collinear). Therefore, with the analysis of $1$, $2$ and $3$
measurements, we have exhaustively addressed the analysis of a single
anchor, whose results depend both on the number of collected
measurements and on their layout on the plane. In light of the results
in Theorem~\ref{thm:single_indistinguishable}, we can define the three
equivalence classes \C1, \C2, and \C3, by introducing the following
notation.
\begin{notation}
  By a set \C1 of measurements, we denote any number of measurements
  collected by the same anchor in the same position $\mathcal{P}$ on
  the plane, provided that $\mathcal{P}$ does not coincide with the
  anchor;\\
  By a set \C2 of measurements, we denote any number of
  \textit{collinear} measurements, not lying on the anchor, collected
  by the same anchor;\\
  By a set \C3 of measurements, we denote any number of measurements
  collected by an anchor, not falling in one of the two cases above,
  i.e. distinct and non-collinear measurement points or with a point
  coinciding with the anchor.
\end{notation}
    \section{Global constructibility with more anchors}
\label{sec:more_anchors}
In this section, we will leverage the results found for a single
anchor to extend the analysis of  indistinguishable trajectories to
the case of multiple anchors. In light of Lemma~\ref{lem:equivalent},
the localisation problem can be solved, i.e. the state of the system
can be determined, if and only if $\mathcal{T}$ is $\ind{1}$.

\subsection{Pathological conditions}
\label{subsec:indistinguishability_large}
While our primary interest is to analyse positive and negative results
for constructibility in the cases in which the available information
is minimal (i.e., small number of anchors), it is useful to discuss
some negative constructibility results that apply to an arbitrarily
large number of anchors and of measurements. This is done in the
following examples.

\begin{figure}[t]
  \centering
  \includegraphics[width=.8\columnwidth]{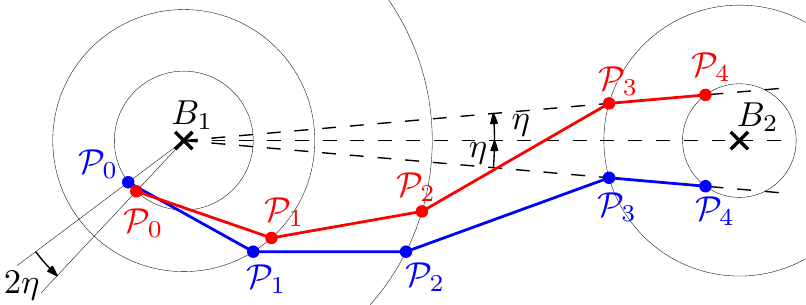}
  \caption{Example~\ref{ex:rotation}. The same trajectory
    $\mathcal{T}$ rotated about the \textit{pivot} anchor $B_1$. When
    $\mathcal{P}_3$, $\mathcal{P}_4$ and $B_1$ are collinear, we
    always have two roto-translations of $\mathcal{T}$ that are
    compliant with the measurements.}
  \label{fig:rotation_example}
\end{figure}

\begin{figure}[t]
  \centering
  \includegraphics[width=.8\columnwidth]{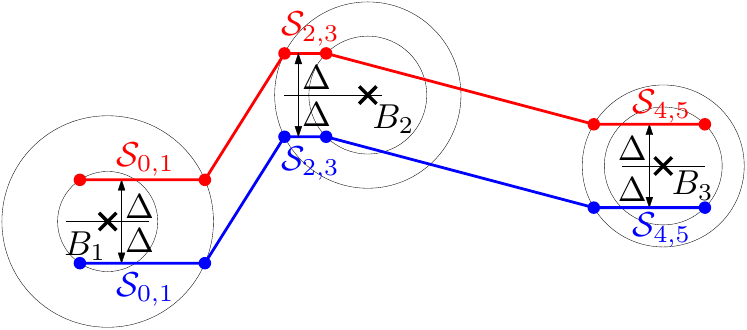}
  \caption{Example~\ref{ex:translation}. The same trajectory
    $\mathcal{T}$ translated by $2\Delta$ orthogonally to
    $\mathcal{S}_{0,1}$. When $\mathcal{S}_{0,1}$, $\mathcal{S}_{2,3}$
    and $\mathcal{S}_{4,5}$ are parallel and have the same distance
    $\Delta$ from the anchor collecting the measurements, we always
    have two translations of $\mathcal{T}$ that are compliant with the
    measurements.}
  \label{fig:translation_example}
\end{figure}

\begin{example}[Rotation]
  \label{ex:rotation}
  Consider the scenario shown in Figure~\ref{fig:rotation_example}. An
  anchor $B_1$ is used to collect a set $\left\{\mathcal{P}_0,
  \mathcal{P}_1, \mathcal{P}_2\right\}$ of \C3 measurements. Consider
  an additional set $\left\{\mathcal{P}_3, \mathcal{P}_4\right\}$ of
  \C2 measurements from a second anchor $B_2$ such that
  $\mathcal{P}_3$ and $\mathcal{P}_4$ are aligned with $B_1$; let
  $\eta$ be the angle between $\mathcal{S}_{3,4}$ and the line
  $\mathcal{B}_{1,2}$ joining the two anchors. If we rotate the whole
  set $\mathcal{T}$ by $2 \eta$ about $B_1$ neither the new readings
  for $\left\{\mathcal{P}_0, \mathcal{P}_1, \mathcal{P}_2\right\}$
  will be affected (Theorem~\ref{thm:single_indistinguishable}), nor
  the new readings for $\left\{\mathcal{P}_3, \mathcal{P}_4\right\}$
  because of the axial symmetry around $\mathcal{B}_{1,2}$. Hence, the
  blue and red trajectories in Figure~\ref{fig:rotation_example} are
  indistinguishable.
\end{example}
\begin{example}[Translation]
  \label{ex:translation}
  Consider the scenario in Figure~\ref{fig:translation_example}. We
  collect a set $\left\{\mathcal{P}_0, \mathcal{P}_1 \right\}$ of \C2
  measurements from anchor $B_1$. Let $\Delta$ be the distance between
  $B_1$ and segment $\mathcal{S}_{0,1}$. By translating the trajectory
  $\mathcal{T}$ by $2 \Delta$ in the direction orthogonal to
  $\mathcal{S}_{0,1}$, we achieve an axial symmetry, which by
  Theorem~\ref{thm:single_indistinguishable} makes the translated
  measurements for $\left\{\mathcal{P}_0, \mathcal{P}_1 \right\}$
  indistinguishable from the previous ones. Consider an additional set
  $\left\{\mathcal{P}_2, \mathcal{P}_3 \right\}$ of \C2 measurements
  from an anchor $B_2$ such that $\mathcal{S}_{2,3}$ is parallel to
  $\mathcal{S}_{0,1}$, and its distance from $B_2$ is $\Delta$. By
  construction, the translation of the trajectory by $2 \Delta$
  generates an axial symmetry on both the anchors. Therefore, the blue
  and red trajectories in Figure~\ref{fig:translation_example} are
  indistinguishable.
\end{example}

\begin{remark}
  Notice that we have presented two examples where each anchor
  collects a set of {\em at least} \C2 of measurements. If one or more
  anchors collect a set \C1 of measurements, the condition for
  indistinguishability simplifies: $\mathcal{S}_{3,4}$ should not be
  necessarily aligned with $B_1$ for Example~\ref{ex:rotation};
  likewise $\mathcal{S}_{2,3}$ should not be necessarily parallel to
  $\mathcal{S}_{0,1}$ for Example~\ref{ex:translation}.
\end{remark}

The construction in the previous examples extends to any number of
anchors, leading to this statement.

\begin{fact}
  \label{claim:N_indistinguishable}
  Given $p$ anchors $B_i = [X_i, Y_i]^\top,\,1,\dots,p$, deployed on a
  plane, there always exists at least one abstract trajectory
  $\mathcal{T}$ for which it is possible to find indistinguishable
  trajectories.
\end{fact}
This fact is the first main result of this paper: given any
configuration of anchors, there is not a sufficiently high number of
measurements and/or anchors such that the system is always
constructible. Luckily, this negative result is limited to specific
pathological trajectories. As discussed next, in the general case it
is possible to overcome this problem.

\subsection{Conditions for unconstructibility}
After discussion some pathological abstract trajectories, which remain
indistinguishable no matter the number of anchors and measurements
taken, we can now shift our focus to the analysis of\emph{ generic
trajectories collecting a small number of measurements} from the
anchors to determine the conditions for unconstructibility. We will
henceforth adopt a special notation to list the number of measurements
collected by each anchor: we will use numbers separated by a ``$+$''
sign, e.g. $3+1$ denotes a \C3 set of measurements collected from the
first anchor and a set of \C1 measurements from the second. Our main
results on necessary and sufficient conditions for
indistinguishability will be constructed analysing this property for
an increasing number of measurements.

\paragraph{$1+1$}
When two anchors are used to collect one measurement each, \emph{a
  single} roto-translation from $\frm V$ to the world frame $\frm W$
  is impossible to construct, which clearly leads to
  unconstructibility. More precisely, given the two measured points
  $\mathcal{P}_0^{(0)}$ and $\mathcal{P}_1^{(0)}$, we can always
  construct an indistinguishable trajectory as follows.  First, we
  rotate $\mathcal{T}$ about $B_1$ of any angle $\phi \in [0,2\pi)$
  (as in the analysis of the single anchor). Assuming that
  $\mathcal{P}_0 ^{(0)}$ has coordinates $[\rho_0, 0]$, its rotated
  version $\mathcal{P}_0$ will have coordinates $\mathcal{P}_0 =
  [\rho_0\cos\phi; \rho_0\sin\phi]^\top$, and  be indistinguishable
  from $\mathcal{P}_0^{(0)}$. In order for the rotated point
  $\mathcal{P}_1$ to be indistinguishable from $\mathcal{P}_1^{(0)}$
  it is sufficient that it lies the intersection between the circle
  centred on $\mathcal{P}_0$ of radius $\|\mathcal{S}_{0,1}\|$ and the
  circle centred on $B_2$ of radius $\rho_1$. Therefore, by assuming
  that $B_1 = [0;0]^\top$, we can find two possible indistinguishable
  points
\begin{equation*}
    \mathcal{P}_1^{(a),(b)} = R_\psi\begin{bmatrix}
        \|\mathcal{S}_{0,1}\| \\ 0
    \end{bmatrix} + R_\phi\begin{bmatrix}
        \rho_0 \\ 0
    \end{bmatrix},
\end{equation*}
where the angle $\psi$ can take one of the following \emph{two} values
(one for each intersection between the two aforementioned circles):
\begin{multline}
    \label{eq:psi_1+1}
    \psi = \phi + \atantwo\left(D\sin\phi, \rho_0-D\cos\phi\right) \\
    \pm\arccos\left(\frac{\rho_1^2-\|\mathcal{S}_{0,1}\|^2-d^2}{2\|\mathcal{S}_{0,1}\|d}\right),
\end{multline}
where $d^2 = D^2+\rho_0^2-2D\rho_0\cos\phi$ is the (unknown) distance
between $\mathcal{P}_0$ and $B_2$, while $D = \|\mathcal{B}_{1,2}\|$
is the distance between the two anchors. $\beta$
in~\eqref{eq:two_measurements_beta} is a particular case of $\psi$,
when the two anchors coincide, i.e. when $D=0$. To summarise,
\emph{this setting generally leads to $\ind{2\times\infty}$
trajectories}. For some particular values of $\phi$, the two circles
become tangent and the two points $\mathcal{P}_1$ coincide. Moreover,
since two pairs of points are involved in this analysis (two anchors
and two measurement points), the trajectories symmetric with respect
to $\mathcal{B}_{1,2}$ are indistinguishable. As a summary we can
state the following:
\begin{case}
  \label{deg_case:straight}
  In the $1+1$ case, generic trajectories are  $\ind{2\times\infty}$
  (hence, indistinguishable). In the degenerate case when
  $\mathcal{S}_{0,1}\subset \mathcal{B}_{1,2}$, the trajectory is
  $\ind 1$.
\end{case}

\begin{figure}[t]
  \centering \includegraphics[width = 0.6\columnwidth]{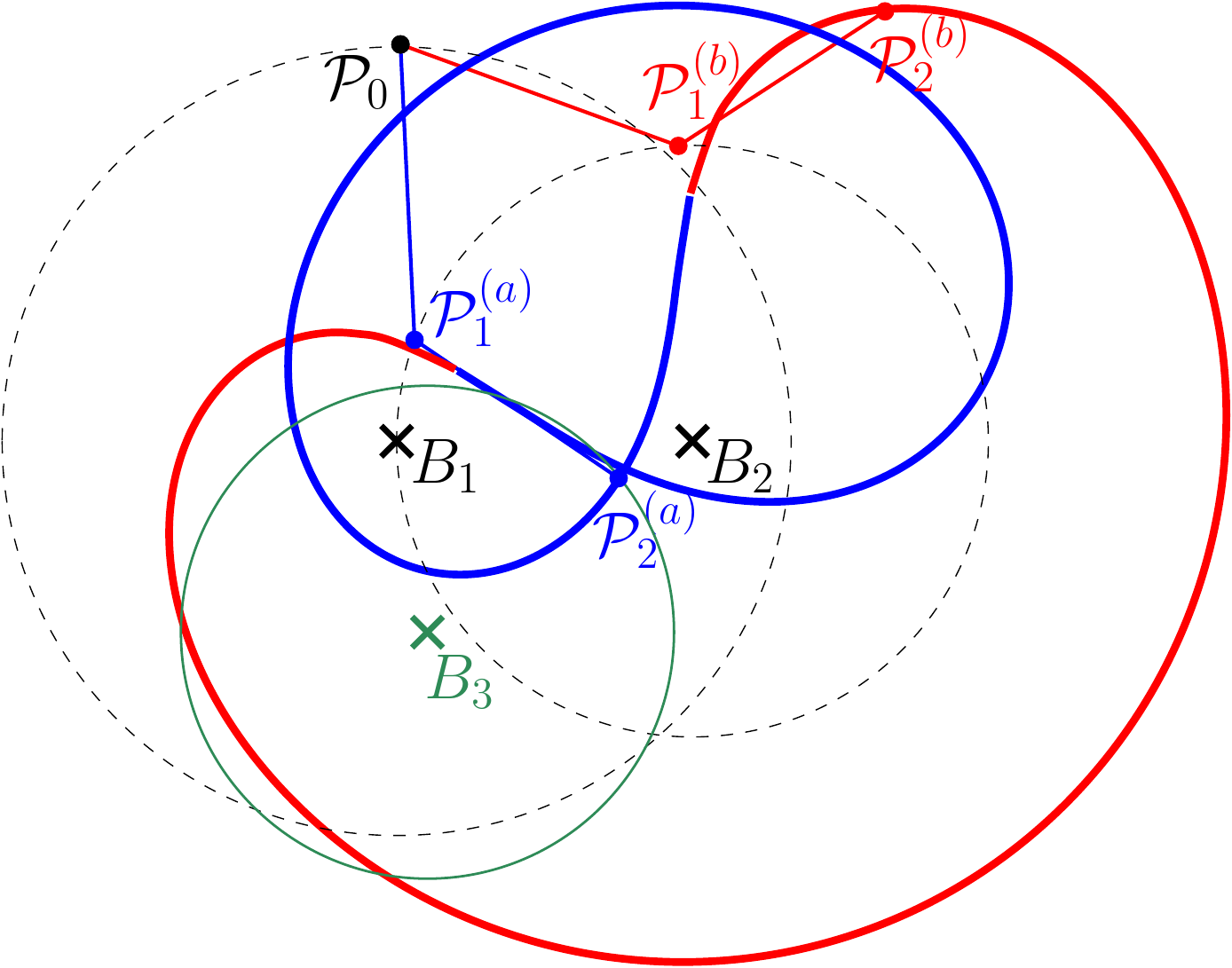}
  \caption{Locus where the third measurement point $\mathcal{P}_2$ can
  lie after the first two measurements $\rho_0$ and $\rho_1$ (dashed
  lines) are collected from the first two anchors. The blue and red
  colours are associated with the two intersections between the
  aforementioned circles. The solid green circle represents the third
  measurement $\rho_2$ collected by $B_3$ in a $1+1+1$ setting. After
  $\rho_2$, the blue and the red trajectories are no more
  indistinguishable ($\mathcal P_2^{(b)}$ does not lie on the green
  circle), but there are still $6$ intersections of the locus with the
  green circle, and thus $\mathcal{T}$ is $\ind{6}$.}
  \label{fig:1+1+1}
\end{figure}

\paragraph{$1+1+1$}
We search for indistinguishable trajectories following the same line
as in the paragraphs above. We start from a trajectory $\mathcal{T}$
characterised by three points $\mathcal{P}_0^{(0)},
\mathcal{P}_1^{(0)}, \mathcal{P}_2^{(0)}$, associated with the
measurements $\rho_0$, $\rho_1$ and $\rho_2$. As for the $1+1$ case,
we rotate the whole trajectory about $B_1$ of an angle $\phi$ and come
up with two potential points $\mathcal{P}_1^{(a)},
\mathcal{P}_1^{(b)}$, which lie on the intersection between a circle
centred on $B_2$ of radius $\rho_1$ and a circle centred onto
$\mathcal{P}_0$ of radius $\mathcal{S}_{0,1}$
(see~\eqref{eq:psi_1+1}). The two points $\mathcal{P}_1^{(a)},
\mathcal{P}_1^{(b)}$ uniquely determine the third potential
measurements $\mathcal{P}_2^{(a)}, \mathcal{P}_2^{(b)}$. By changing
$\phi$, the points $\mathcal{P}_2^{(a)}, \mathcal{P}_2^{(b)}$ generate
the locus shown in Figure~\ref{fig:1+1+1}. The locus is parametrised
by the angle $\phi$ (and thus its dimension is $1$), it is defined
only when $\bigl|\|\mathcal{S}_{0,1}\| - d\bigr| < \rho_1 < \|\mathcal
S_{0,1}\| + d$, and it is continuous and differentiable on its domain.
Indistinguishability arises when the locus intersects the circle
centred on $B_3$ of radius $\rho_2$ in more than one point, i.e. when
$\mathcal{P}_2^{(a)}$ and $\mathcal{P}_2^{(b)}$ have the same sensor
readings.

As a consequence, the indistinguishable third point $\mathcal{P}_2$
has \emph{at most} $8$ different locations. Indeed, by defining $C =
\cos\phi$ and $S = \sin\phi$, we take the differences
$\rho_2^2-\rho_0^2$ and $\rho_1^2-\rho_0^2$ and obtain linear
equations in the unknowns $C$ and $S$, thus yielding a unique solution
$(\bar C, \bar S)$. Then we impose the constraints $\bar C^2+\bar S^2
= 1$ and $x_0^2+y_0^2 = \rho_0^2$, being polynomials with degree $4$
and $2$ in the unknowns $x_0$ and $y_0$, respectively. By Bezout's
theorem, the maximum number of real solutions $(x_0,y_0)$ to this set
of equations is the product between the degrees of the two
polynomials, i.e. $8$.

Finally, also for the $1+1+1$ case, we can have a degenerate case, as
detailed next.
\begin{case}
  \label{deg_case:1+1+1}
  In the $1+1+1$ case, generic trajectories are $\ind{\bar n},\,\bar n
  \le 8$, and thus indistinguishable. In the degenerate case when the
  locus is tangent to the circle centred on $B_3$ of radius $\rho_2$
  in one point, the trajectory $\mathcal{T}$ is $\ind 1$.
\end{case}

\paragraph{$2+1$}
We address the case $2+1$ following the same procedure as in the
previous case. From~\eqref{eq:two_points_distance} we can compute the
distances $d_2^{(a)}$ and $d_2^{(b)}$ of the points
$\mathcal{P}_2^{(a)}$ and $\mathcal{P}_2^{(b)}$ from $B_1$, and the
locus in Figure~\ref{fig:1+1+1} degenerates to two circles centred on
$B_1$. Each of the two circles has two intersections as long as
$\underline{d} < \|\mathcal{P}_2^\star - B_1\| < \overline{d}$ holds
true, where $\underline{d} = |D-\rho_2|$ and $\overline{d} =
D+\rho_2$, and $\mathcal{P}_2^\star$ denotes any of the two points
$\mathcal{P}_2^{(a)}$ and $\mathcal{P}_2^{(b)}$. Hence, there are
\textit{at most} four indistinguishable trajectories. This particular
condition allows us to compute the number of indistinguishable
trajectories. Moreover, we can introduce a degenerate case for this
setting.
\begin{case}
  \label{deg_case:2+1}
  The $2+1$ setting is a particular case of the $1+1+1$ setting, since
  the locus in Figure~\ref{fig:1+1+1} is a pair of circles centred on
  the first anchor. Generic trajectories are $\ind 4$, while in the
  degenerate cases when $\min\{d_2^{(a)},d_2^{(b)}\} = \|B_2-B_1\| -
  \rho_2$ or $\max\{d_2^{(a)},d_2^{(b)}\} = \|B_2-B_1\| + \rho_2$, the
  circles are tangent in a point lying on $\mathcal{B}_{1,2}$, and the
  trajectory is $\ind{1}$.
\end{case}
\paragraph{$3+1$}
With the same \textit{rationale}, we reconstruct the distance $d_3$
between $\mathcal{P}_3$ and $B_1$. Therefore, $\mathcal{P}_3$ lies on
the intersection between the circle centred on $B_1$ of radius $d_3$
and the circle centred on $B_2$ of radius $\rho_3$, thus yielding two
intersections with the following degenerate case.
\begin{case}
\label{deg_case:3+1}
In the $3+1$ case, generic trajectories are $\ind 2$. When $d_3 = D
\pm \rho_3$, the two circles are tangent in a point lying on  the line
connecting $B_1$ and $B_2$. In this degenerate case, the two
intersections, i.e. the two indistinguishable trajectories, collapse
on each other, thus achieving $\ind 1$.
\end{case}
With the definition of these four degenerate cases, keeping in mind
that the role of the two anchors can be switched (i.e., $3 + 1$ is
equivalent to $1 + 3$), and considering that any number $N_m$ of
measurements can be at most a \C3 set of measurements, we can now
state the following theorem.

\begin{theorem}
\label{thm:negative_result}
Given a trajectory $\mathcal{T}$ with $N_m$ measurement points, the
system is unconstructible when two anchors collecting $N_m - 1$ and
$1$ measurement, respectively, are involved, or when $N_m \le 3$,
unless at least one among the degenerate cases in
Case~\ref{deg_case:straight}, \ref{deg_case:1+1+1}, \ref{deg_case:2+1}
and~\ref{deg_case:3+1} occurs.
\end{theorem}

We have shown a set of settings where constructibility is never
achieved, or it is achieved only for some particular shapes of
$\mathcal{T}$ and layouts of the anchors. These setting are summarised
in red in Figure~\ref{fig:summing_up}.

\subsection{Conditions for constructibility}
\label{subsec:CondConstr}
We now consider all the other cases and search for constructibility
conditions, keeping in mind that trajectory indistinguishability may
arise when pathological trajectories are selected, as stated in
Section~\ref{subsec:indistinguishability_large}.

\begin{figure}[t]
  \centering
  \includegraphics[width=0.7\columnwidth]{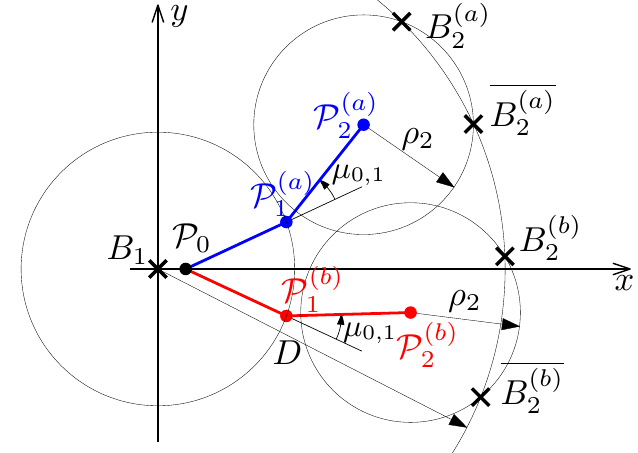}
  \caption{New reference frame showing the setting $2+1$. The blue and
    red lines represent the two trajectories $\mathcal{T}^{(a)}$,
    $\mathcal{T}^{(b)}$. Each of them has a circle centred on their
    last point $\mathcal{P}_2$, hence yielding an overall number of
    $4$ intersections (i.e. possible positions of $B_2$) with the
    circle centred on $B_1$ of radius $D$.}
  \label{fig:two_plus_one}
\end{figure}

\paragraph{$2+2$}
For this analysis, we will define a new reference frame, which will
simplify the forthcoming discussion. To this aim, we consider
Equation~\eqref{eq:two_points_distance}, with $\phi=0$. This way, we
know the position of the first anchor $B_1$, lying on the origin, and
of the two trajectories $\mathcal{T}^{(a)}$ and $\mathcal{T}^{(b)}$,
as shown in Figure~\ref{fig:two_plus_one}.
In this particular reference frame, the $4$ indistinguishable
trajectories arising in the setting $2+1$ correspond to $4$ positions
of the second anchor $B_2$. For each pair of indistinguishable
trajectories, we want to analyse how a further measurement collected
by the second anchor preserves or solves the ambiguity. At first, we
notice that ambiguities may arise between two trajectories rotated
both about $B_1$ and about $\mathcal{P}_0$, i.e. $\mathcal{T}^{(a)}$
and $\mathcal{T}^{(b)}$, or between two trajectories only rotated
about the anchor $B_1$, i.e. $\mathcal{T}_1$ and $\mathcal{T}_2$, and
thus the analysis will be divided into two parts, one for each pair of
trajectories.

\textbf{Rotation about anchor:} Given the two measurement points
$\mathcal{P}_2 = [x_2,y_2]^\top$ and $\mathcal{P}_3 = [x_3,y_3]^\top$,
we want to find the position of the anchor $B_2$ satisfying the
following equations:
\begin{equation}
    \label{eq:fourth_measure}
    \arraycolsep=1pt
    \left\{
    \begin{array}{ccccc}
         X_2^2      &+&  Y_2^2      &=& D^2,\\
        (X_2-x_2)^2 &+& (Y_2-y_2)^2 &=& \rho_2^2,\\
        (X_2-x_3)^2 &+& (Y_2-y_3)^2 &=& \rho_3^2.
    \end{array}\right.
\end{equation}
To this end, we take the difference of the last two equations with
respect to the first and get to these linear equations in the unknowns
$X_2,Y_2$
\begin{equation}
    \label{eq:M_fourth_measure}
    M B_2 = \begin{bmatrix}
        x_2 & y_2\\
        x_3 & y_3
    \end{bmatrix} \begin{bmatrix}
        X_2\\Y_2
    \end{bmatrix} = \frac{1}{2}\begin{bmatrix}
        D^2 - \rho_2^2 + x_2^2 + y_2^2\\
        D^2 - \rho_3^2 + x_3^2 + y_3^2
    \end{bmatrix}.
\end{equation}
To find a unique solution for $B_2$, we need a nonsingular matrix $M$,
whose determinant is
\begin{equation}
  \label{eq:M_singular_beta}
  \det M = x_3 y_2 - x_2 y_3 .
\end{equation}
Therefore, $B_2$ has a unique solution, i.e. \emph{there exists no
pair of indistinguishable trajectories rotated about the first anchor,
if $B_1$, $\mathcal{P}_2$ and $\mathcal{P}_3$ are not aligned}. Hence,
to guarantee $\ind 1$, $\mathcal{P}_3$ cannot lie on the two lines
joining $B_1$ and $\mathcal{P}_2^{(a)}$, and joining $B_1$ and
$\mathcal{P}_2^{(b)}$, which are available in the reference frame
$\frm V$. From a geometric point of view, we may reformulate the
problem as finding the position of $B_2$ by using three ranging
measurements. Indeed, by using trilateration this problem has a unique
solution if the three ranging measurement points are non-collinear.
This result is perfectly in line with the scenario proposed in
Example~\ref{ex:rotation}.

\textbf{Rotation about anchor and initial point:} Given two points
$\mathcal{P}_2^{(a)}$, $\mathcal{P}_3^{(a)}$, we can derive
$\mathcal{P}_2^{(b)}$, $\mathcal{P}_3^{(b)}$ as
$\mathcal{P}^{(b)}_\star = R_\zeta (\mathcal{P}^{(a)}_\star -
\mathcal{P}_0) + \mathcal{P}_0$, where $\zeta = -2\delta$ and $\delta$
defined in~\eqref{eq:two_measurements_beta}, and the subscript $\star$
is either $2$ or $3$. With these two pairs, we want to find the
positions of two anchors $B_2^{(a)}$, $B_2^{(b)}$ satisfying the set
of equations~\eqref{eq:M_fourth_measure}, for both $\mathcal{T}^{(a)}$
and $\mathcal{T}^{(b)}$. With the same rationale followed previously,
we take the differences
\begin{equation}
\label{eq:two_two_differences}
\begin{split}
    \|\mathcal{P}_2^{(a)} - B_2^{(a)} \|^2 - \|B_2^{(a)}\|^2 =
    \|\mathcal{P}_2^{(b)} - B_2^{(b)} \|^2 - \|B_2^{(b)}\|^2 \\ 
    \|\mathcal{P}_3^{(a)} - B_2^{(a)} \|^2 - \|B_2^{(a)}\|^2 =
    \|\mathcal{P}_3^{(b)} - B_2^{(b)} \|^2 - \|B_2^{(b)}\|^2
\end{split}
\end{equation}
As in the previous case, we get to two linear equations as $M
[X_2^{(a)},Y_2^{(a)}]^\top = h$, with the same $M$ as
in~\eqref{eq:M_singular_beta}. With this result, given one of the two
feasible $B_2^{(b)}$ obtained in the case $2+1$, we find a unique
anchor $B_2^{(a)}$ satisfying the differences of the distances,
provided that $\mathcal{P}_2^{(a)}$, $\mathcal{P}_3^{(a)}$ and $B_1$
are not aligned, as before. Since $D$ is the distance from $B_1$ to
$B_2$ and $B_1$ is in the origin of the reference frame, we now add
the constraint $\|B_2\| = D$, i.e. $\|B_2^{(a)}\| - \|B_2^{(b)}\| =
0$, with $B_2^{(a)}$ obtained as the unique solution
of~\eqref{eq:two_two_differences}. Therefore, we have a quadratic
equation in the coordinates of $\mathcal{P}_3^{(a)}$ in the form
$[x_3;y_3;1]^\top\, Q \,[x_3;y_3;1]=0$, where the matrix of the
quadratic equation $Q$, representing a conic section, is
\begin{equation*}
    Q = \begin{bmatrix}
        S & b \\ b^\top & c
    \end{bmatrix},
\end{equation*}
where $S\in \mathbb{R}^{2\times 2}$, $b\in \mathbb{R}^{2}$ and $c\in
\mathbb{R}$, and where its invariants characterise the conic. In
particular, the centre of the conic is $O = -S^{-1} b =
\mathcal{P}_2^{(a)}$, while $\det Q = 0$, and thus this is a
degenerate conic with centre $\mathcal{P}_2^{(a)}$. To identify its
shape, we analyse the determinant of the submatrix $S$, that yields
\begin{equation*}
    \det S = -\rho_0^2 (X_2^{(b)} - \mathcal{X}_2^{(b)})^2
    (\|\mathcal{P}_2^{(a)}\|^2 - {\underline{d}}^2)
    ({ \overline{d}}^2 - \|\mathcal{P}_2^{(a)}\|^2),
\end{equation*}
where $\mathcal{X}_2^{(b)} =
X_2^{(b)}\cos\zeta-\rho_0(\cos\zeta-1)+Y_2^{(b)}\sin\zeta$ is the $x$
coordinate of the point obtained by rotating $B_2^{(b)}$ about
$\mathcal{P}_0$ by $-\zeta$. The condition $\underline{d} <
\|\mathcal{P}_2\| < \overline{d}$ guarantees that the product of the
last two terms is always positive, while the intermediate term is
always nonpositive, and it is $0$ when the points $\mathcal{P}_0$,
$\mathcal{P}_1^{(b)}$ and $B_2^{(b)}$ are collinear. This situation is
the mirrored version of the situation analysed above, where the two
measurement points collected from $B_2$ and $B_1$ were aligned, and
thus there exists no points $\mathcal{P}_3^{(a)}$ that can recover
$\ind 1$, as in Example~\ref{ex:rotation}. In fact, matrix $Q$ is in
this case the $0$ matrix, i.e. a conic describing the whole $X_w
\times Y_w$ motion plane. When this unfortunate situation does not
occur, the determinant is negative, hence the conic described by $Q$
is a degenerate hyperbole, i.e. two lines intersecting in
$\mathcal{P}_2^{(a)}$ and thus, for each of the two positions
$B_2^{(b)}$ arising from the setting $2+1$, we find two critical
lines.

In conclusion, \emph{we have two critical directions for
$\mathcal{P}_3^{(a)}$ arising from the first situation and four from
the second}, and thus there exists $6$ lines in $\frm V$, intersecting
in $\mathcal{P}_2^{(a)}$, where $\mathcal{P}_3^{(a)}$ should not lie
onto to ensure that the trajectory is $\ind 1$.

\paragraph{$2+1+1$}
With respect to the previous case, we here collect the same number of
measurements, but we distribute them among $3$ anchors. One can follow
the same procedure as before, obtaining more convoluted expressions
leading to the same result with a more complex geometrical
interpretation. However, as in the previous case, we can conclude
that, in the reference frame $\frm V$ there are at most $6$ lines
where $\mathcal{P}_3$ should not lie onto to achieve a $\ind 1$
problem.  Indeed, given the (at most) four indistinguishable
trajectories arising in the setting $2+1$, we can find the four
possible positions of the third anchor in $\frm V$. We can compute the
distances between a given fourth position $\mathcal{P}_{V,3}$ and each
of the four ``virtual'' anchors $B_{V,3}^{(i)},\,i=1,\dots,4$. Two
among these distances coincide, i.e.
$\|B_{V,3}^{(i)}-\mathcal{P}_{V,3}\| =
\|B_{V,3}^{(j)}-\mathcal{P}_{V,3}\|,\,i\neq j$, if and only if
$\mathcal{P}_{V,3}$ lies on the axis of the segment having as vertexes
a pair of the ``virtual'' anchors themselves. If $\mathcal{P}_{V,3}$
lies on one of these $6$ critical lines, then the system is $\ind{2}$,
while when $\mathcal{P}_{V,3}$ does not lie on any of these lines,
$\mathcal{T}$ is $\ind{1}$.

\paragraph{$3+2$}
The analysis carried out in this section is a particular case of the
setting $2+2$. Indeed, by collecting \C3 measurements from the first
anchor, we can discard one of the two indistinguishable trajectories
$\mathcal{T}^{(a)}$ and $\mathcal{T}^{(b)}$. Therefore,
indistinguishability can be obtained only by rotation about the first
anchor, coming up with a set of equations as
in~\eqref{eq:fourth_measure}, with the proper modifications on the
subscript to account for the additional point sensed by the first
anchor. We get to the same conclusion as
in~\eqref{eq:M_singular_beta}, i.e. that the trajectories are
indistinguishable only if $B_1$, $\mathcal{P}_3$ and $\mathcal{P}_4$
are aligned.

\paragraph{$3+1+1$}
With the same \textit{rationale} as in the analysis of the $3+2$
setting based on the $2+2$ scenario, we can adapt here the analysis of
the setting $2+1+1$. After the first $4$ measurements ($3+1$), there
exist two indistinguishable trajectories and thus, in the reference
frame $\frm V$, there is a line (obtained with the same procedure
presented in the setting $2+1+1$) where $\mathcal{P}_4$ should not lie
to solve this ambiguity.

\paragraph{$1+1+1+1$}
As discussed in the analysis of the setting $1+1+1$, after three
measurements, there is a finite number $\overline n < 8$ of
indistinguishable trajectories. As in the previous cases, for each of
the (at most) $28$ pairs there is a line where $\mathcal{P}_3$ should
not lie to achieve a $\ind 1$ trajectory.

With these findings, we can state the following theorem.
\begin{theorem}
  \label{thm:positive result}
  Given a trajectory $\mathcal{T}$ and $N_m$ measurement points, the
  system is constructible when $N_m \ge 4$ and each anchor collects at
  most $N_m-2$ measurements, unless the last point of $\mathcal{T}$
  lies on one of the indistinguishability line identified in the
  analysis.
\end{theorem}

\begin{figure}[t]
  \includegraphics[width=\columnwidth]{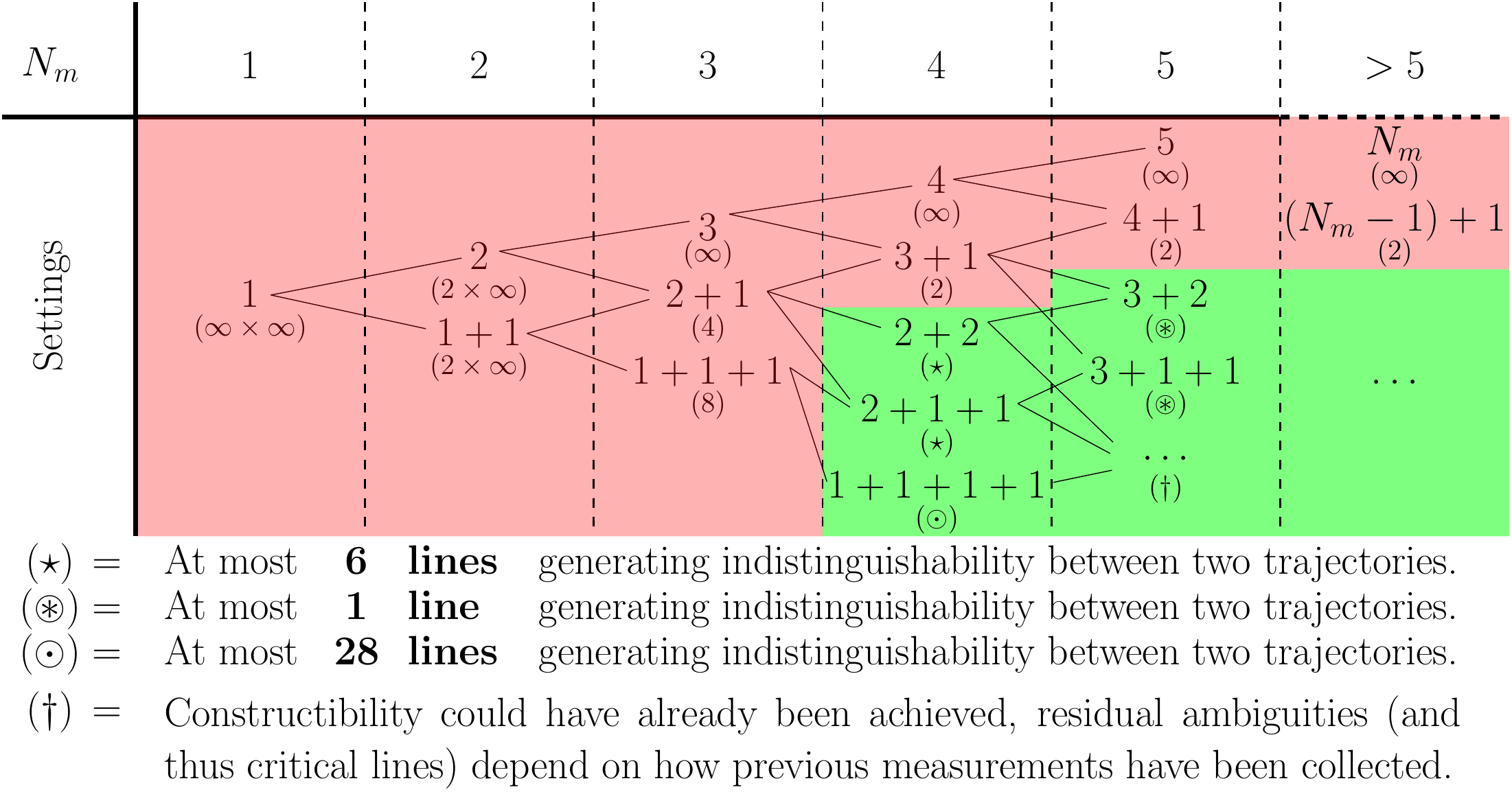}
  \caption{Summarising picture subsuming the taxonomy derived in this
    paper as a function of the overall number of measurements and of
    their distribution among the different anchors. The number in
    brackets denotes the number of indistinguishable trajectories. The
    red part is referred to Theorem~\ref{thm:negative_result}, while
    the green part is associated with the results obtained in
    Theorem~\ref{thm:positive result}.}
  \label{fig:summing_up}
\end{figure}

Using the necessary and sufficient conditions to attain
constructibility identified previously, the final taxonomy of
Figure~\ref{fig:summing_up} can be derived, where the area highlighted
in red subsumes the results of Theorem~\ref{thm:negative_result},
while the part highlighted in green is referred to
Theorem~\ref{thm:positive result}.

\begin{remark}
  Apparently, there is a duality between the conditions for
  constructibility in Theorem~\ref{thm:negative_result}
  and~\ref{thm:positive result}. However, from a practical view point,
  in the latter case, the vehicle can compute numerically in $\frm V$
  the ``critical'' lines before collecting the last measurement, plan
  its last manoeuvre to avoid such lines and achieve $\ind 1$. On the
  other hand, in the former scenario, the vehicle is not able to plan
  its trajectory to fall into the degenerate
  cases~\ref{deg_case:straight}, \ref{deg_case:1+1+1},
  \ref{deg_case:2+1}, \ref{deg_case:3+1}, since they are detected once
  all the measurements are collected.
\end{remark}


\begin{remark}
\label{rem:mapping}
We now reverse the perspective, by considering the problem of mapping,
the dual problem with respect to localisation. In this case, we want
to find the position of the anchors $B_i = [X_i; Y_i]^\top$ in the
reference frame $\frm V$, where the trajectory $\mathcal{T}$ of the
vehicle is known. Although the two problems are dual, there are
remarkable differences in the analysis. Indeed, in the localisation
problem, we have used both the shape of the trajectory $\mathcal{T}$
and the layout of the anchors $B_i$, while here we have no information
on the layout of the anchors on the plane. Thus, the overall mapping
problem boils down to an independent mapping problem for each anchor,
which leads to the classic trilateration
problem~\cite{palopoli2020global}.
\end{remark}
    \section{Local Constructibility}
\label{sec:local_constructibility}
In the previous sections, we have focused our analysis on the concept
of \emph{indistinguishability} of trajectories. In this section we
shift our focus on \emph{local} (weak) constructibility properties of
the system, and we show that when some \emph{a priori} information is
known, e.g. an estimate of the initial state of the system, a lower
number of measurements is needed to reconstruct the state of the
vehicle. As expected, some results devised in the previous sections
are inherited for this analysis, i.e. weak constructibility is
\emph{usually} ensured when the trajectory $\mathcal{T}$ is $\ind{1}$,
while other settings (e.g. the degenerated
cases~\ref{deg_case:straight}, \ref{deg_case:1+1+1},
\ref{deg_case:2+1}, and~\ref{deg_case:3+1}) yield pathological
conditions from a local perspective. To analyse \emph{weak}
constructibility, we build the \textit{Constructibility Gramian} of
the system, which depends explicitly on the system dynamics.
Therefore, we consider a unicycle--like vehicle collecting ranging
measurements. The state $q$ of the system includes its Cartesian
coordinates $x$,$y$ and its heading $\theta$ with respect to a
reference axis, and has the following dynamics:
\begin{equation}
    \label{eq:unicycle}
    \dot x = v\cos\theta,\qquad\dot y = v\sin\theta,\qquad\dot\theta=\omega,
\end{equation}
where the forward velocity $v$ and the angular velocity $\omega$ of
the vehicle are the control inputs $u$. Being the unicycle kinematic
model a differentially flat system with its position as flat
output~\cite{fliess1995flatness}, it satisfies
Property~\ref{property:dynamical_systems} and
Assumption~\ref{assum:initial}.

\subsection{Constructibility Gramian}
\begin{figure}[t]
  \centering
  \includegraphics[width=0.6\columnwidth]{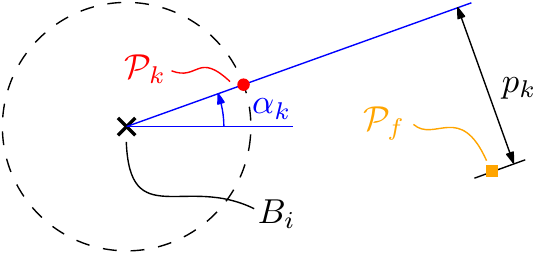}
  \caption{The vehicle is located in the point $\mathcal{P}_k$, and it
    is sensed by the anchor $B_i$. The quantities affecting the term
    $g_k$ of the CG only depend on the angle $\alpha_k$ of the
    measurement and not on the collected distance $\rho_k$.}
  \label{fig:single_measurement_CG}
\end{figure}

We build the Constructibility Gramian (CG) for the unicycle kinematic
model~\eqref{eq:unicycle} subject to intermittent ranging
measurements~\eqref{eq:DT_output_equation}. The CG is an $n\times n$
matrix, where $n$ is the size of the state of the system, that may be
used as a tool to check for nonlinear constructibility, i.e. it
describes how difficult it is to reconstruct the final state of the
system given the controls and the measurements over a time window
$[t_0,t_f]$. In particular, the reciprocal of its smallest singular
value (or equivalently of its smallest eigenvalue, since the CG is
symmetric and positive semidefinite), quantifies how measurement noise
affects the estimate of the final state, and if it is equal to $0$,
i.e. if the CG is singular, then the system is \emph{weakly
unconstructible} (see~\cite{krener2009measures}). The CG is defined
either for continuous- or discrete-time systems, and thus we have to
extend its definition to a generic system with continuous-time
dynamics and intermittent (discrete-time) measurement
output~\eqref{eq:CT_DT}. For a continuous-time system
\begin{equation*}
    \dot q(t) = f\bigl(q(t),u(t)\bigr),\qquad
         z(t) = h\bigl(q(t)     \bigr),
\end{equation*}
the Constructibility Gramian $G_C(t_0,t_f)$ is defined as
\begin{equation}
    \label{eq:CT_CG}
    G_C(t_0,t_f) = \int_{t_0}^{t_f} \Phi^\top(\tau,t_f) H^\top(\tau)
    W_C(\tau) H(\tau) \Phi(\tau,t_f)\,d\tau,
\end{equation}
where $H(t) = \left(\frac{\partial h}{\partial q}\right)
\Bigr|_{q=q(t)}$ is the Jacobian of the measurement evaluated at the
current time $t$, and $W_C(t)$ is a weighing matrix, which accounts
for heterogeneous measurement units, different uncertainties among
sensors, or for nonlinear effects, such as bounded sensing range. The
\textit{sensitivity matrix} $\Phi(t,t_f) = \frac{\partial
q(t)}{\partial q_f}$ represents how small perturbations in the final
condition of the system affect the state at the current time $t$, and
is the unique solution to the final value problem:
\begin{equation*}
    \dot \Phi (t,t_f) = F(t) \Phi(t,t_f),\qquad
    \Phi(t_f,t_f)     = \mathbf{I}_n,
\end{equation*}
where $F(t) = \left(\frac{\partial f}{\partial q} \right)
\Bigr|_{q=q(t)}$ is the linearised dynamics of the system and
$\mathbf{I}_n$ is the $n\times n$ identity matrix. To model
intermittent measurements, we design $W_C$ such that
\begin{equation*}
  W_C(t_k) = \left\{ \begin{array}{ll}
                       \delta_D\diag(\underline{e}_i) & \mbox{ if $B_i$ is measured at $t_k$,} \\
                       \mathbf 0 & \mbox{ otherwise.}
    \end{array} \right.
\end{equation*}
where $\mathbf 0$ denotes the null matrix, the $\diag(w)$ operator
indicates a diagonal matrix having the entries of the vector $w$ as
diagonal entries, $\underline{e}_i$ is the unitary vector aligned with
the $i$-th axis, thus assuming that the sensors have homogeneous
uncertainty, and $\delta_D$ is the Dirac delta. This way, with the
same idea as in the previous section, we can disregard the dynamics of
the system and the trajectory followed by the vehicle and focus on a
finite number of points $\mathcal{P}_k$ where the measurements are
collected. However, the definition of the CG explicitly contains the
final state $q_f$ and, by computations carried out hereafter, it
depends on the final position $\mathcal{P}_f = [x_f,y_f]^\top$,
reached by the vehicle at time $t = t_f$. Therefore, the CG
in~\eqref{eq:CT_CG} may be rewritten in the following simplified
expression
\begin{equation*}
    G_C(t_0,t_f) = \sum\limits_{k=0}^{N_m-1} g(t_k,t_f),
\end{equation*}
where $N_m$ is the overall number of measurements, and the
contribution $g(t_k,t_f)$ of the $k$-th measurement, denoted as $g_k$
in the following, is computed with its definition in~\eqref{eq:CT_CG}
\begin{equation}
    \label{eq:v_CG}
    g_k = \gamma_k\gamma_k^\top, \qquad
    \gamma_k = [\cos \alpha_k, \sin\alpha_k, p_k]^\top,
\end{equation}
with $\alpha_k = \atantwo(y_k-Y_{i},x_k - X_{i})$, while $p_k$ is the
distance of the final point from the line passing through the measured
anchor $B_i$ and having slope $\alpha_k$, i.e.
\begin{equation*}
    p_k = \frac{(x_f-x_k)(Y_{i}-y_k) - (y_f-y_k)(X_{i}-x_k)}
        {\sqrt{(x_k-X_{i})^2+(y_k-Y_{i})^2}} .
\end{equation*}
Notice that $p_k$ is not dependent on the collected measurement, i.e.
the distance $\rho_k$. Figure~\ref{fig:single_measurement_CG} shows
the relevant parameters defining $g_k$.
By construction, $g_k$ is an $n \times n$ matrix with rank $1$, whose
column space is $\gamma_k$. We will leverage considerations on the
rank of sum of matrices, relying on alignment among null and column
spaces of the contributions $g_k$. To formally define the relationship
between \emph{weak constructibility} properties and rank of the CG, we
report here the following theorem, adapted
from~\cite{powel2015empirical}, whose proof is skipped for brevity.
\begin{theorem}[\cite{powel2015empirical}]
Given system~\eqref{eq:CT_DT}, if there exists a sequence of control
inputs $u(t),\,t\in [t_0,t_f]$, i.e. a shape of $\mathcal{T}$, such
that $CG(t_0,t_f)$ is full rank, then system~\eqref{eq:CT_DT} is
\emph{weakly constructible}.
\end{theorem}
As in the previous sections, we analyse the settings with increasing
number of measurement and anchors.

\subsection{Single anchor}
With a single anchor collecting measurements, the analysis of the CG
trivially leads to the same conclusions drawn in
Theorem~\ref{thm:single_indistinguishable}, i.e. the trajectory can
rotate about the anchor without modifying the sensor readings.

\paragraph{One measurement}
When a single measurement is collected, in $\mathcal P_0 =
[x_0,y_0]^\top$, the Constructibility Gramian is simply computed as $G
= \gamma_0 \gamma_0^\top$, where $\gamma_0$ is defined
in~\eqref{eq:v_CG} and thus, by construction, the CG has rank $1$. Its
null space, i.e. the unconstructible subspace, is a two-dimensional
vector subspace with the vectors $[-(y_f-Y_1); x_f-X_1; 1]$ and
$[-(y_f-y_0); x_f-x_0; 1]$ as basis. The two vectors defining the
unconstructible subspace are tangent to the circles passing through
$P_f$ and centred on $B_1$ and on $\mathcal P_0$, respectively. This
result is compliant with Theorem~\ref{thm:single_indistinguishable},
hence highlighting the same constructibility properties.

\begin{figure}[t]
  \centering
  \begin{tabular}{c|c}
    \includegraphics[width=0.4\columnwidth]{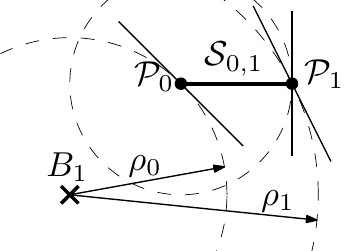} &
                                                            \includegraphics[width=0.4\columnwidth]{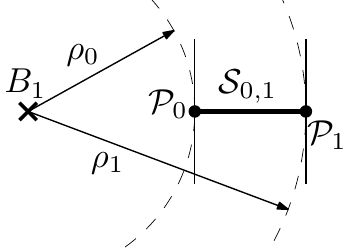}
                                                            \\
    (a) & (b)
  \end{tabular}
  \caption{A pair of measurements collected by anchor $B_1$. (a) The
    three points are not aligned, the rank of the CG is $2$ and the
    two tangents of the circles passing through $\mathcal{P}_1$ are
    not aligned. (b) The particular situation where $B_1$,
    $\mathcal{P}_0$ and $\mathcal{P}_1$ are aligned occurs, the two
    circles passing through $\mathcal{P}_1$ share the same tangent,
    hence making the CG rank deficient.}
  \label{fig:CG_2}
\end{figure}

\paragraph{Two measurements}
The second measurement collected by the anchor generates an overall
Constructibility Gramian $G = \gamma_0 \gamma_0^\top + \gamma_1
\gamma_1^\top$, having at most rank $2$, since it is the sum of two
rank $1$ matrices. Since the column space of $g_0$ is $\gamma_0$ by
construction, we can analytically derive the conditions on
$\mathcal{P}_1$ such that the Gramian has still rank $1$, by solving
$\gamma_1 = \ell \gamma_0$, with $\ell \in \mathbb{R}$, yielding
$\mathcal{P}_1 = \ell \mathcal{P}_0 + (1-\ell) B_1$, i.e.
$\mathcal{P}_0$, $\mathcal{P}_1$ and $B_1$ are collinear, occurring
whenever the vehicle is moving on the diameter of the circle centred
on the anchor. From an analytical viewpoint, this result is not
surprising, since $\mathcal{P}_0$ and $\mathcal{P}_1$ share the same
angle $\alpha$, and thus $\gamma_0 = \gamma_1$. This result may be
interpreted by keeping in mind that we are dealing with local
properties, i.e. we are regarding rotations as (small) translations
along the tangent of the circle centred on the rotation pole. In this
particular scenario, the rotation about $B_1$ and about
$\mathcal{P}_0$ share the same tangent, orthogonal to
$\mathcal{S}_{0,1}$, and thus, only in this setting, also a rotation
about $\mathcal{P}_0$ is allowed. The two situations with rank $2$ and
rank $1$ are represented in Figure~\ref{fig:CG_2}.

\begin{remark}
  \label{rem:two_measurements_diameter_local}
  This result has a strong connection with the scenario described in
  Remark~\ref{rem:two_measurements_diameter_global}, where
  indistinguishability was avoided, while preventing weak
  constructibility. In this situation, the angle $\delta$ defined
  in~\eqref{eq:two_measurements_beta} is equal to $0$ and the two
  points $\mathcal{P}_1^{(a)}$ and $\mathcal{P}_1^{(b)}$ collapse on
  each other.
\end{remark}

Any further measurement collected by the first anchor, beyond the
second, is not informative from a local point of view (provided that
the two measurement points are not collinear with the anchor itself).
Indeed, the vector $[-(y_f-Y_1); x_f-X_1; 1]^\top$, associated with
the rotation of the trajectory about the anchor, does not depend on
the sensed measurement point. Hence, a single anchor always generates
a singular Gramian, independently on the number of measurements
collected.

\subsection{Two anchors}
As in Section~\ref{sec:more_anchors}, we now consider a higher number
of anchors and an increasing number of measurements distributed among
them. Since the maximum number of informative measurements collected
by an anchor is $2$, we will analyse the settings $1+1$, $2+1$ and
$2+2$ hereafter.

\paragraph{$1+1$}
Since we are summing two rank-$1$ matrices, we can already state
upfront that the CG will be singular. However, with the same rationale
of the previous case, we look for the condition on $\mathcal{P}_1$
such that the Gramian has rank $1$, i.e. when $\gamma_1 = \ell
\gamma_0$, thus getting from the first two equations
\begin{equation}
    \label{eq:1_plus_1_P1}
    \mathcal P_1 = \ell (\mathcal P_0 - B_1) + B_2.
\end{equation}
By plugging this definition of $\mathcal{P}_1$ into the third
equation, which reads as $p_1 = \ell p_0$, we get a linear equation in
$y_0$, yielding
\begin{equation*}
    y_0 = \frac{Y_1-Y_2}{X_1-X_2} x_0 + \frac{X_1Y_2-X_2Y_1}{X_1-X_2},
\end{equation*}
i.e. $\mathcal{P}_0$, $B_1$ and $B_2$ collinear. By plugging this
result in~\eqref{eq:1_plus_1_P1}, we get that also $\mathcal{P}_1$
lies on the same line. Thus, as soon as the four points are not
collinear, the rank of the Gramian is $2$.

\begin{remark}
This condition has already been discussed in the degenerate
Case~\ref{deg_case:straight}, where the circles passing through
$\mathcal{P}_1$ and centred on $\mathcal{P}_0$ and $B_2$,
respectively, share the same tangent. In this situation, two
indistinguishable trajectories coincide, thus achieving $\ind 1$ and
preventing weak constructibility.
\end{remark}

\paragraph{$2+1$}
Without loss of generality, let us consider two measurements in
$\mathcal{P}_0$ and $\mathcal{P}_1$ from $B_1$, and one in
$\mathcal{P}_2$ from $B_2$
\begin{equation*}
    \arraycolsep=1pt
    \ker (\gamma_0\gamma_0^\top + \gamma_1\gamma_1^\top) =
    [-(y_f-Y_1),\quad x_f-X_1,\quad1]^\top,
\end{equation*}
while the column space of $\gamma_2^\top \gamma_2$ is $\gamma_2$
itself. The CG has rank $2$ when these two vectors are orthogonal,
i.e. their inner product $\langle \ker (\gamma_0\gamma_0^\top +
\gamma_1\gamma_1^\top), \gamma_2 \rangle$ is $0$, hence when
\begin{equation*}
    X_1Y_2 -X_2Y_1 -X_1y_2 +Y_1x_2 +X_2y_2 -Y_2x_2 = 0,
\end{equation*}
i.e. when $B_1$, $B_2$ and $\mathcal{P}_2$ are aligned. To give a
geometrical interpretation, we need to refer to the results obtained
in the case $2+1$ in Section~\ref{sec:more_anchors} (see
Figure~\ref{fig:two_plus_one}). From a local perspective we have some
knowledge on the initial state of the system, i.e. we can a priori
distinguish whether the vehicle is travelling along the trajectory
$\mathcal{T}^{(a)}$ or $\mathcal{T}^{(b)}$ in
Figure~\ref{fig:two_plus_one}, and thus we can compute the distance of
$\mathcal{P}_2$ from the anchor $B_1$. Therefore, we know that
$\mathcal{P}_2$ lies on the intersection between two circles centred
in $B_1$ and $B_2$ respectively. Whenever $\mathcal{P}_2$ lies on the
line connecting the two anchors, these two circles intersect in a
single point, and therefore they share the same tangent direction,
with the same conclusions as in the $1+1$ setting, described in
Figure~\ref{fig:CG_2}. Notice that, in this case, we do not have a
perfect duality with the degenerated case~\ref{deg_case:2+1}, since we
can discard a priori one of the two circles in
Figure~\ref{fig:two_plus_one}.

\paragraph{$2+2$}
We consider two pairs of measurement points being not collinear with
the anchor collecting their distance. In this scenario, the CG is
singular as long as the $1$-dimensional null spaces of the Gramians
$G_1$, $G_2$, associated with each anchor, are aligned. The condition
$\ker (G_2) = \ell \ker (G_1)$ reads
\begin{equation*}
  -y_f + Y_2 = -\ell y_f + \ell Y_1, \quad
    x_f - X_2 =  \ell x_f - \ell X_1, \quad
          1  =  \ell,
\end{equation*}
hence yielding $B_1 = B_2$, which is impossible by assumption of
distinct anchors. Therefore, when two anchors collect a pair of
measurements each, the system is weakly constructible as far as the
pair of measurement points and the anchor collecting their distances
are not collinear.

\subsection{Three anchors}
With three anchors, we only consider the scenario $1+1+1$, which is
expected to yield results similar to the case $2+1$. We build the
column spaces $\gamma_k,\,k=0,1,2$, of the three contributions to the
CG. The overall Gramian is full rank as soon as the three column
spaces are linearly independent, and this conditions may be checked by
computing the determinant of $W = [\gamma_0,\gamma_1,\gamma_2]$,
yielding $\det W = a x_2 + b y_2 + c$, where $a$, $b$, $c$ are three
parameters depending on the coordinates of the three anchors $B_1$,
$B_2$, $B_3$, and of the two measurement points $\mathcal{P}_0$ and
$\mathcal{P}_1$. Hence, $\det W = 0$ describes a line on the
$X_w\times Y_w$ plane, passing through the anchor $B_3$. Therefore,
the Gramian is singular as soon as $\mathcal{P}_2$ lies on this line,
with known analytical form. From a geometric point of view, this line
has a similar interpretation to the one obtained in the scenario
$2+1$. Indeed, by combining the rotation of $\mathcal{P}_1$ about
$B_1$ and of $\mathcal{P}_2$ about $B_2$ such that $\mathcal{S}_{1,2}$
maintains the same length, the (tangent to the) resulting motion of
the third point $\mathcal{P}_3$ is tangent to the circle centred on
$B_3$ and passing through $\mathcal{P}_3$ itself. From a global
perspective, in this situation two intersections between two 1D
geometrical varieties coincide (see
Remarks~\ref{rem:two_measurements_diameter_global}
and~\ref{rem:two_measurements_diameter_local}), but there are
guarantees on its uniqueness.

By the conclusions drawn so far, the main result on \emph{weak
constructibility} is stated in the following Theorem.
\begin{theorem}
  \label{thm:local_constructibility}
  Given at least three measurements, distributed among at least $2$
  anchors, the trajectory $\mathcal{T}$ is \textit{weakly
  constructible}, unless the last point of $\mathcal{T}$ lies on one
  of the critical lines identified in the analysis.
\end{theorem}
    \section{Conclusions}
\label{sec:conclusions}
We have proposed an analysis of indistinguishability by abstracting
the manoeuvres executed by the vehicle and the measurements collected
by the ranging sensors, in order to have an easy geometric
interpretation of the results obtained in this analysis for the
localisation problem. Then, we have proposed a local constructibility
analysis based on the Constructibility Gramian, showing analysis and
differences between the local and global analysis. In the near future,
we plan to develop control strategies considering the anchors as
moving agents of a multiagent system, ensuring both global
constructibility and desired levels of local constructibility that can
be quantified through some norm of the CG, in the spirit of the active
sensing. Before using the obtained results in single- or multi--agent
trajectory planning, a deeper analysis on the effect of actuation
uncertainties and measurement noises would be beneficial to the
synthesis of planning algorithms.
    \bibliographystyle{IEEEtran}
    \bibliography{references}

\begin{thebibliography}{10}
\providecommand{\url}[1]{#1}
\csname url@samestyle\endcsname
\providecommand{\newblock}{\relax}
\providecommand{\bibinfo}[2]{#2}
\providecommand{\BIBentrySTDinterwordspacing}{\spaceskip=0pt\relax}
\providecommand{\BIBentryALTinterwordstretchfactor}{4}
\providecommand{\BIBentryALTinterwordspacing}{\spaceskip=\fontdimen2\font plus
\BIBentryALTinterwordstretchfactor\fontdimen3\font minus
  \fontdimen4\font\relax}
\providecommand{\BIBforeignlanguage}[2]{{%
\expandafter\ifx\csname l@#1\endcsname\relax
\typeout{** WARNING: IEEEtran.bst: No hyphenation pattern has been}%
\typeout{** loaded for the language `#1'. Using the pattern for}%
\typeout{** the default language instead.}%
\else
\language=\csname l@#1\endcsname
\fi
#2}}
\providecommand{\BIBdecl}{\relax}
\BIBdecl

\bibitem{MagnagoCPPF19iros}
V.~Magnago, P.~Corbal\'an, G.~Picco, L.~Palopoli, and D.~Fontanelli, ``{Robot
  Localization via Odometry-assisted Ultra-wideband Ranging with Stochastic
  Guarantees},'' in \emph{Proc. IEEE/RSJ International Conference on
  Intelligent Robots and System (IROS)}.\hskip 1em plus 0.5em minus 0.4em\relax
  Macao, China: IEEE, Nov. 2019, pp. 1607--1613.

\bibitem{Chen2013survey}
P.~Chen, Y.~B. Xu, L.~Chen, and Z.~A. Deng, ``{Survey of WLAN Fingerprinting
  Positioning System},'' \emph{Applied Mechanics and Materials}, vol. 380, pp.
  2499--2505, Aug. 2013.

\bibitem{NazemzadehMFMP15}
P.~Nazemzadeh, F.~Moro, D.~Fontanelli, D.~Macii, and L.~Palopoli, ``{Indoor
  Positioning of a Robotic Walking Assistant for Large Public Environments},''
  \emph{IEEE Trans. on Instrumentation and Measurement}, vol.~64, no.~11, pp.
  2965--2976, Nov 2015.

\bibitem{Cheok10}
N.~Cheok, M.~Radovnikovich, P.~Vempaty, G.~Hudas, J.~Overholt, and P.~Fleck,
  ``{UWB} tracking of mobile robots,'' in \emph{Proc. IEEE International
  Symposium on Personal Indoor and Mobile Radio Communications (PIMRC)},
  Instanbul, Turkey, Sep. 2010, pp. 2615--2620.

\bibitem{Gallant16}
M.~J. Gallant and J.~A. Marshall, ``{Two-Dimensional Axis Mapping Using
  LiDAR},'' \emph{IEEE Trans. on Robotics}, vol.~32, no.~1, pp. 150--160, Feb.
  2016.

\bibitem{PosnerB20}
D.~Barnes and I.~Posner, ``{Under the Radar: Learning to Predict Robust
  Keypoints for Odometry Estimation and Metric Localisation in Radar},'' in
  \emph{2020 IEEE International Conference on Robotics and Automation (ICRA)},
  2020, pp. 9484--9490.

\bibitem{belo2013complete}
F.~A. Belo, P.~Salaris, D.~Fontanelli, and A.~Bicchi, ``A complete
  observability analysis of the planar bearing localization and mapping for
  visual servoing with known camera velocities,'' \emph{International Journal
  of Advanced Robotic Systems}, vol.~10, no.~4, p. 197, 2013.

\bibitem{sert2012localizability}
H.~Sert, W.~Perruquetti, A.~Kokosy, X.~Jin, and J.~Palos, ``Localizability of
  unicycle mobiles robots: An algebraic point of view,'' in \emph{2012 IEEE/RSJ
  International Conference on Intelligent Robots and Systems}.\hskip 1em plus
  0.5em minus 0.4em\relax IEEE, 2012, pp. 223--228.

\bibitem{mariottini2005vision}
G.~L. Mariottini, G.~Pappas, D.~Prattichizzo, and K.~Daniilidis, ``Vision-based
  localization of leader-follower formations,'' in \emph{Proceedings of the
  44th IEEE Conference on Decision and Control}.\hskip 1em plus 0.5em minus
  0.4em\relax IEEE, 2005, pp. 635--640.

\bibitem{wu2020robust}
Y.~Wu, Y.~Li, W.~Li, H.~Li, and R.~Lu, ``{Robust LiDAR-based localization
  scheme for unmanned ground vehicle via multisensor fusion},'' \emph{IEEE
  Transactions on Neural Networks and Learning Systems}, vol.~32, no.~12, pp.
  5633--5643, 2020.

\bibitem{delaune2021range}
J.~Delaune, D.~S. Bayard, and R.~Brockers, ``Range-visual-inertial odometry:
  Scale observability without excitation,'' \emph{IEEE Robotics and Automation
  Letters}, vol.~6, no.~2, pp. 2421--2428, 2021.

\bibitem{martinelli2017unicycle}
A.~Martinelli, ``The unicycle in presence of a single disturbance:
  Observability properties,'' in \emph{2017 Proceedings of the Conference on
  Control and its Applications}.\hskip 1em plus 0.5em minus 0.4em\relax SIAM,
  2017, pp. 62--69.

\bibitem{de2017underwater}
D.~De~Palma, F.~Arrichiello, G.~Parlangeli, and G.~Indiveri, ``Underwater
  localization using single beacon measurements: Observability analysis for a
  double integrator system,'' \emph{Ocean Engineering}, vol. 142, pp. 650--665,
  2017.

\bibitem{quenzer2014observability}
J.~D. Quenzer and K.~A. Morgansen, ``Observability based control in range-only
  underwater vehicle localization,'' in \emph{2014 American control
  conference}.\hskip 1em plus 0.5em minus 0.4em\relax IEEE, 2014, pp.
  4702--4707.

\bibitem{fernando2021toward}
E.~Fernando, O.~De~Silva, G.~K. Mann, and R.~Gosine, ``{Toward Developing an
  Indoor Localization System for MAVs Using Two or Three RF Range Anchors: An
  Observability Based Approach},'' \emph{IEEE Sensors Journal}, vol.~22, no.~6,
  pp. 5173--5187, 2021.

\bibitem{MagnagoPBTMNMF20tim}
V.~Magnago, L.~Palopoli, A.~Buffi, B.~Tellini, A.~Motroni, P.~Nepa, D.~Macii,
  and D.~Fontanelli, ``{Ranging-free UHF-RFID Robot Positioning through Phase
  Measurements of Passive Tags},'' \emph{IEEE Trans. on Instrumentation and
  Measurement}, vol.~69, no.~5, pp. 2408--2418, May 2020.

\bibitem{fontanelli2021uncertainty}
D.~Fontanelli, F.~Shamsfakhr, D.~Macii, and L.~Palopoli, ``An
  uncertainty-driven and observability-based state estimator for nonholonomic
  robots,'' \emph{IEEE Transactions on Instrumentation and Measurement},
  vol.~70, pp. 1--12, 2021.

\bibitem{araki2019range}
B.~Araki, I.~Gilitschenski, T.~Ogata, A.~Wallar, W.~Schwarting, Z.~Choudhury,
  S.~Karaman, and D.~Rus, ``Range-based cooperative localization with nonlinear
  observability analysis,'' in \emph{2019 IEEE Intelligent Transportation
  Systems Conference (ITSC)}.\hskip 1em plus 0.5em minus 0.4em\relax IEEE,
  2019, pp. 1864--1870.

\bibitem{hung2020range}
N.~T. Hung and A.~M. Pascoal, ``Range-based navigation and target localization:
  Observability analysis and guidelines for motion planning,''
  \emph{IFAC-PapersOnLine}, vol.~53, no.~2, pp. 14\,620--14\,627, 2020.

\bibitem{coleman2021observability}
D.~Coleman, S.~D. Bopardikar, and X.~Tan, ``Observability-aware target tracking
  with range only measurement,'' in \emph{2021 American Control Conference
  (ACC)}.\hskip 1em plus 0.5em minus 0.4em\relax IEEE, 2021, pp. 4217--4224.

\bibitem{mandic2016mobile}
F.~Mandi{\'c}, N.~Mi{\v{s}}kovi{\'c}, N.~Palomeras, M.~Carreras, and
  G.~Vallicrosa, ``Mobile beacon control algorithm that ensures observability
  in single range navigation,'' \emph{IFAC-PapersOnLine}, vol.~49, no.~23, pp.
  48--53, 2016.

\bibitem{cedervall2007nonlinear}
S.~Cedervall and X.~Hu, ``Nonlinear observers for unicycle robots with range
  sensors,'' \emph{IEEE transactions on automatic control}, vol.~52, no.~7, pp.
  1325--1329, 2007.

\bibitem{salaris2019online}
P.~Salaris, M.~Cognetti, R.~Spica, and P.~R. Giordano, ``{Online Optimal
  Perception-Aware Trajectory Generation},'' \emph{IEEE Transactions on
  Robotics}, vol.~35, no.~6, pp. 1307--1322, 2019.

\bibitem{napolitano2021gramian}
O.~Napolitano, D.~Fontanelli, L.~Pallottino, and P.~Salaris,
  ``{Information-aware Lyapunov-based MPC in a feedback-feedforward control
  strategy for autonomous robots},'' \emph{IEEE Robotics and Automation
  Letters}, vol.~7, no.~2, pp. 4765--4772, April 2022.

\bibitem{bayat2015range}
M.~Bayat, N.~Crasta, A.~P. Aguiar, and A.~M. Pascoal, ``{Range-based underwater
  vehicle localization in the presence of unknown ocean currents: Theory and
  experiments},'' \emph{IEEE Transactions on control systems technology},
  vol.~24, no.~1, pp. 122--139, 2015.

\bibitem{palopoli2020global}
L.~Palopoli and D.~Fontanelli, ``{Global observability analysis of a
  nonholonomic robot using range sensors},'' in \emph{2020 European Control
  Conference (ECC)}.\hskip 1em plus 0.5em minus 0.4em\relax IEEE, 2020, pp.
  1300--1305.

\bibitem{riz2022local}
F.~Riz, L.~Palopoli, and D.~Fontanelli, ``{On Local/Global Constructibility for
  Mobile Robots using Bounded Range Measurements},'' \emph{IEEE Control Systems
  Letters}, vol.~6, pp. 3038--3043, 2022.

\bibitem{fliess1995flatness}
M.~Fliess, J.~L{\'e}vine, P.~Martin, and P.~Rouchon, ``Flatness and defect of
  non-linear systems: introductory theory and examples,'' \emph{International
  journal of control}, vol.~61, no.~6, pp. 1327--1361, 1995.

\bibitem{krener2009measures}
A.~J. Krener and K.~Ide, ``Measures of unobservability,'' in \emph{Proceedings
  of the 48h IEEE Conference on Decision and Control (CDC) held jointly with
  2009 28th Chinese Control Conference}.\hskip 1em plus 0.5em minus 0.4em\relax
  IEEE, 2009, pp. 6401--6406.

\bibitem{powel2015empirical}
N.~D. Powel and K.~A. Morgansen, ``Empirical observability gramian rank
  condition for weak observability of nonlinear systems with control,'' in
  \emph{2015 54th IEEE Conference on Decision and Control (CDC)}.\hskip 1em
  plus 0.5em minus 0.4em\relax IEEE, 2015, pp. 6342--6348.

\end{thebibliography}

\end{document}